\documentclass[default]{sn-jnl}

\jyear{2022}%

\theoremstyle{thmstyleone}%
\usepackage{lscape}
\usepackage{multirow}
\usepackage{graphicx} 
\usepackage{lscape}
\theoremstyle{thmstyletwo}%

\theoremstyle{thmstylethree}%
\usepackage{graphicx}
\usepackage{hyperref}
\usepackage{url}
\begin{document}

\title[Article Title]{
Machine learning can guide experimental approaches for protein digestibility estimations} 


\author*[1]{\fnm{Sara} \sur{Malvar}}\email{saramalvar@microsoft.com}
\author[2]{\fnm{Anvita} \sur{Bhagavathula}}
\author[1]{\fnm{Maria Angels de Luis} \sur{Balaguer}}
\author[1]{\fnm{Swati} \sur{Sharma}}
\author[1]{\fnm{Ranveer} \sur{Chandra}}

\affil*[1]{\orgname{Microsoft Research}, \orgaddress{\street{Microsoft Building 99, 15010 NE 36th St 2}, \city{Redmond}, \postcode{09805}, \state{Wa}, \country{USA}}}
\affil[2]{\orgname{Brown University}, \orgaddress{\city{Providence}, \postcode{02912}, \state{RI}, \country{USA}}}


\abstract{Food protein digestibility and bioavailability are critical aspects in addressing human nutritional demands, particularly when seeking sustainable alternatives to animal-based proteins. In this study, we propose a machine learning approach to predict the true ileal digestibility coefficient of food items. The model makes use of a unique curated dataset that combines nutritional information from different foods with FASTA sequences of some of their protein families. We extracted the biochemical properties of the proteins and combined these properties with embeddings from a Transformer-based protein Language Model (pLM). In addition, we used SHAP to identify features that contribute most to the model prediction and provide interpretability. This first AI-based model for predicting food protein digestibility has an accuracy of 90\% compared to existing experimental techniques. With this accuracy, our model can eliminate the need for lengthy \textit{in-vivo} or \textit{in-vitro} experiments, making the process of creating new foods faster, cheaper, and more ethical. }






\maketitle

\section{Introduction}\label{sec1}

The increase in the world population and the need for more nutritious food will increase the demand for animal-based foods by nearly 70\%, particularly from ruminant meat \cite{world_bank_world_2007}. Closing this food gap will intensify the pressure on land usage and contribute to an increase in greenhouse gas (GHG) emissions \cite{henchion_future_2017}. 
Due to such environmental concerns, transitioning towards more sustainable diets and exploring alternative protein sources have been at the forefront of 21st century research \cite{nadathur_sustainable_2017}. 

Although plant proteins have already been associated with lowering the risk of type 2 diabetes, cardiovascular diseases, hypertension, obesity, metabolic syndrome and all-cause mortality in prospective cohort studies \cite{yokoyama_vegetarian_2014,satija_plant-based_2016, dinu_vegetarian_2017,kim_plantbased_2019}, there is still a long way to go in terms of texture, flavor, and bioavailability. 
A common concern among consumers is recognizing which foods are good protein sources and whether proteins from plants are of as good quality as those from animal-derived foods. The interaction between animal and plant proteins is being comprehensively investigated to develop balanced mixtures of animal and vegetable proteins. These studies involve determining and understanding protein structure-function relations, optimizing the use of the components of the product, improving the quality, reducing costs, and finding new protein applications.

Protein quality is defined by the essential amino acid composition of the protein, as well as the bioavailability and digestibility of its constituent amino acids. Over the last few decades, tremendous effort has been made to develop methodologies for measuring protein quality. 
The Food and Agriculture Organization of the United Nations (FAO) currently advises evaluating protein quality in human diets using Digestible Indispensable Amino Acid Score (DIAAS) values, which are measured through human or pig ileal digesta \cite{fao_dietary_2013}. Unlike older methods such as Protein Digestibility Corrected Amino Acid Score (PDCAAS) values, DIAAS values are obtained by multiplying the digestibility of each essential AA by its concentration in the protein and then comparing the results to a scoring system \cite{rutherfurd_protein_2015}. This digestion process can also be simulated \textit{in-vitro}. In this case, the digestibility is measured by monitoring the amount of soluble proteins, changes in digesta pH (reflecting protein hydrolysis), or the level of increase in the nitrogen not incorporated in protein structures (i.e., nonprotein nitrogen) \cite{karlund_harnessing_2020}. Although \textit{in-vitro} analysis can be cheaper and easier than the methods to predict the outcome of \textit{in-vivo}
digestibility, the complexity of \textit{in-vivo} digestibility has not been realized fully in an \textit{in-vitro} model \cite{bohn_correlation_2018}. In addition, working with animals can be time consuming, costly and one could face ethical barriers. 

Currently, an important goal for food manufacturers is to replicate the flavor profiles, texture, and bioavailability of a meat replacement. This approach has traditionally been quite repetitious, with hundreds of alternative protein source formulations eventually narrowed down to the most comparable profile. In this work, we aim to provide a solution to accelerate this process at a reduced cost through the use of an AI model for protein digestibility prediction. 
Our contributions are as follows:
\begin{enumerate}
        \item \textbf{Dataset creation} We combine nutritional information such as dietary fiber content, fat, and vitamins of various foods with the FASTA sequences of their protein families. We use these sequences for a feature extraction step that aims to identify protein structural characteristics. The extracted structural features and the nutritional information formed the basis of our dataset used to predict the DIAAS value.  The feature vector for each food consists of their structural and nutritional characteristics, amino acid composition and some categorical variables related to the food type. To our knowledge, this is the first attempt to associate these disparate features and use them in a machine learning (ML) model.
    \item \textbf{Interpretability} 
   We select the features with higher predictive power, and, consequently, facilitate the model's interpretability. Selecting these features enhanced the model's performance due to dimensionality reduction. Exploring model interpretability is an important step to guide new experiments and new food design processes, as it gives insight on how protein engineering can be used to change protein's structural information to increase a food's digestibility. Similarly, it can guide how we can add/remove salts, decrease dietary fiber content and manipulate the nutritional information to achieve higher bioavailability. In addition to that, we were able to confirm some important proved correlations between food nutrients and protein digestibility.
    \item \textbf{ML modeling} We developed an alternative ML-based approach to predict the DIAAS score that avoids \textit{in-vivo} and \textit{in-vitro} testing, is cost-effective, and fast. With a 90\% accuracy compared to experimental approaches, our model can be used to guide experimental approaches by testing new protein sources in a faster, cheaper, and ethical manner. Our ML model targets true ileal digestibility coefficient of each indispensable amino acid, which allows us to calculate DIAAS for each food item, considering the reference values.
\end{enumerate}

\section{Results}\label{sec2}
\subsection{Dataset curation}

To be able to train a model that can predict digestibility, it is necessary to obtain training data. To our knowledge, there was no public ground truth data available at this point, especially for DIAAS, as it has been proposed more recently. Thus, a complete database was developed for this analysis using the following original data:

\begin{itemize}
    \item First, a database containing ileal amino acid digestibility values for more than 180 food items along with the associated DIAAS values in cases where adequate information was reported to derive this score \cite{molly_muleya_ileal_2021} was used. This data was obtained from over 30 references presented in supplementary table \ref{references}. This feature is defined as our target variable.
    \item Nutritional information for foods such as fat, potassium and sodium content was obtained from data from the FoodData Central of the United States Department of Agriculture (USDA) \cite{us_department_of_agriculture_usda_agricultural_research_service_nutrient_data_usda_nodate} and Food Data from the National Food Institute in the Technical University of Denmark \cite{FRIDA}. For each food item in the database, the features presented in supplementary tables \ref{table1} and \ref{table2} were also added. 
    \item In order to capture additional protein features, the FASTA sequence of at least two of the most abundant proteins from different protein families of each food item were added to the dataset. Then, their physio-chemical and biochemical properties were generated using Protlearn \cite{dorfer_protlearn_nodate} and added to the database. These features are shown in section \ref{features_protlearn} of the Supplementary Material.
\end{itemize}


\subsection{Baseline models}

First, 12 baseline regressors were trained on all 1671 features of 189 food items. The R$^2$ values and RMSE of the 12 models is shown in section \ref{regressors} of the Supplementary Material. The 3 top-performing models were based on boosting and bagging techniques applied to decision trees: LightGBM, XGBoost and Random Forest. The target variable was defined as the true ileal digestibility coefficient of each indispensable amino acid, as further detailed in Methods. As reported in section \ref{regressors} of the Supplementary Material, we found that the Random Forest, XGBoost, and LightGBM models predicted the ileal digestibility coefficient with a validation $R^{2}$ of 0.83, 0.87, and 0.88 respectively.

\subsection{Feature selection and SHAP values}

\begin{figure}[h!]
  \includegraphics[width=\linewidth]{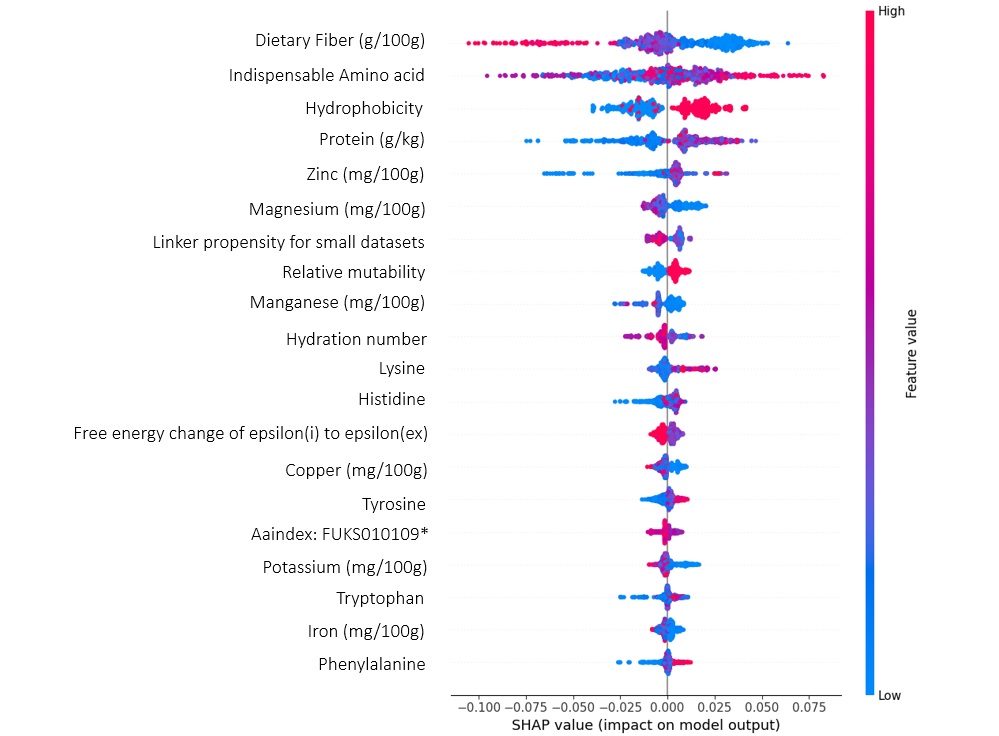}
  \caption{The local explanation summary of the model represents a set of beeswarm 
plots, where each dot corresponds to an individual sample in the study. The dot’s position on the x axis shows the impact that feature has on the model’s 
prediction for that person. When multiple dots land at the same x position, they pile up to show density.}
  \label{fig2}
\end{figure}

The second step of many ML methods is the feature extraction, the aim of which is to get the most effective features from the obtained raw segment. This problem was approached as a feature selection or dimensionality reduction method, which is pervasive in all domains of application of machine learning and data mining. To understand the importance of the features in the model and provide some explainability, a sensitivity based method called SHAP \cite{guyon_advances_2017} was used for feature importance evaluation \cite{lundberg_unified_2017} (see Methods). 
First, we determined the importance of the features using SHAP. Then, we used Principal Component Analysis \cite{noauthor_principal_2002} and determined that the top 20 features explained above 95\% of the model variance. Therefore, we selected these features for our model, as shown in figure \ref{fig2}. The color represents the feature value, where red indicates high and blue indicates low. It is noticeable that some of the SHAP-selected features presented as important are known in the literature for their direct effect on digestibility. For instance, Mongeau \cite{mongeau_relationship_1989} shows that the true protein digestibility in rats was negatively correlated with the total food fiber level. Likewise, the type of limiting amino acid is directly associated with the true ileal digestibility of that amino acid. Salts like magnesium and manganese are also associated with a lower rate of protein digestibility, disrupting the enzymatic process \cite{kaspchak_effect_2019,geboes_magnesium_2002}. 

Protein digestibility is also found to be positively correlated with surface hydrophobicity of proteins through a study on heat-induced protein unfolding \cite{tang_heat-induced_2022}. Hydrophobic groups are normally buried within protein structure. However, the unfolding of proteins exposes hydrophobic residues to the surface, increasing overall surface hydrophobicity and providing easier access for digestive enzymes to hydrolyze the protein \cite{tang_heat-induced_2022}.  Our model found other features associated with surface hydrophobicity, such as hydration number and linker propensity, to also confirm this correlation. For example, the hydration number indicates how likely a water molecule is to bind to a solute as compared to other water molecules \cite{zavitsas_opinions_2016}, implying that the hydration number is inversely linked to hydrophobicity and therefore to protein digestibility too \cite{sagawa_are_2014}. Similarly, linker propensity, or a tendency of a protein to have high numbers of linkers, can be linked to hydrophilicity since preferred linker amino acids have been established to be mostly hydrophilic \cite{sagawa_are_2014}. 

Several minerals have also been shown to impact digestibility in accordance with our Shapley value results. For instance, Pallauf and Kirchgessner demonstrate that zinc deficiency has a negative effect on nutrient digestibility in weaned male rats \cite{pallauf_einflus_1976}. Salts containing potassium, such as Potassium thiocyanate, were shown to be trypsin inhibitors, which are in turn known to reduce the digestion and absorption of dietary proteins \cite{shikimi_modes_1979} \cite{aviles-gaxiola_inactivation_2018}. Additionally, specific amino acids have been established as having an impact on protein digestibility. Amino acids such as lysine and histidine have been demonstrated to have a positive impact on food intake and metabolism \cite{baruffol_l-lysine_2014} \cite{moro_histidine_2020}. For example, L-lysine delayed intestinal transit in rats \cite{baruffol_l-lysine_2014}, potentially increasing ileal protein absorption by increasing the time the protein resided in the small intestine \cite{zhao_protein_1996}. Other amino acids, such as Tryptophan and phenylalanine, have been shown to modulate gut microbiota and the secretion of digestive enzymes \cite{taleb_tryptophan_2019} \cite{guo_phenylalanine_2018}, likely increasing the amount of protein absorption.

\subsection{Protein embeddings from language model}

It is known that protein structure helps in understanding function. To be able to understand how the structure of proteins modify the digestibility of proteins, we used a language model approach. Here, we assume that the protein sequence is a series of tokens, or characters, such as a text corpus. Protein sequences are inherently similar to natural languages: amino acids arrange in a multitude of combinations to form structures that carry function, the same way as letters form words and sentences, which carry meaning 
\cite{ferruz_controllable_2022}. 

\begin{figure}
  \includegraphics[scale=0.9]{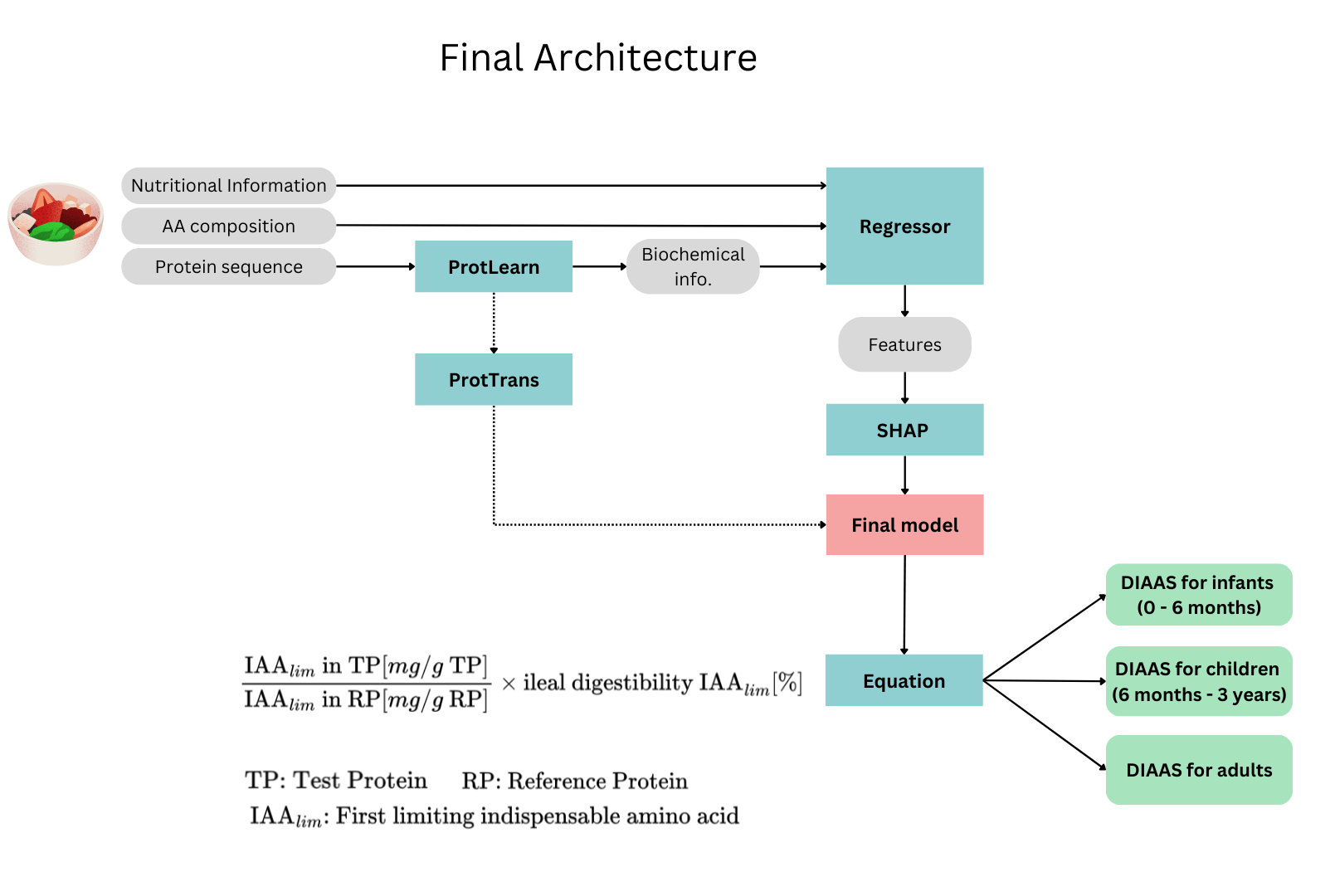}
  \caption{Final architecture of the proposed method. For each food item, nutritional information, amino acid composition and protein sequences of at least two families are required. The algorithm includes two feature extraction steps: biochemical information is extracted using Protlearn \cite{dorfer_protlearn_nodate} and embeddings are extracted using ProtT5 \cite{elnaggar_prottrans_2022}. The baseline regressors are trained and two feature selection approaches are compared: feature importance from LGBM and SHAP \cite{guyon_advances_2017}. The true ileal digestibility for each indispensible amino acid and food item is predicted and the DIAAS value is calculated for three different categories: infants, children and adults.}
  \label{fig1}
\end{figure}

To make the model more robust, we extracted embeddings from the protein sequences using ProtTrans, a language model trained with UniRef50 and BFD100 \cite{elnaggar_prottrans_2022} and added these embeddings to the final dataset. The final architecture of our method is shown in figure \ref{fig1}.

\subsection{Model results}
When creating this unique machine learning model, we used several concepts that contribute to the overall model performance. In the ablation study, we try to identify the influence of each of these innovations separately. Details of the ablation study can be found on section \ref{ablation} of the Supplementary Material. 

\begin{figure}[h!]
\centering
  \includegraphics[scale=1.3]{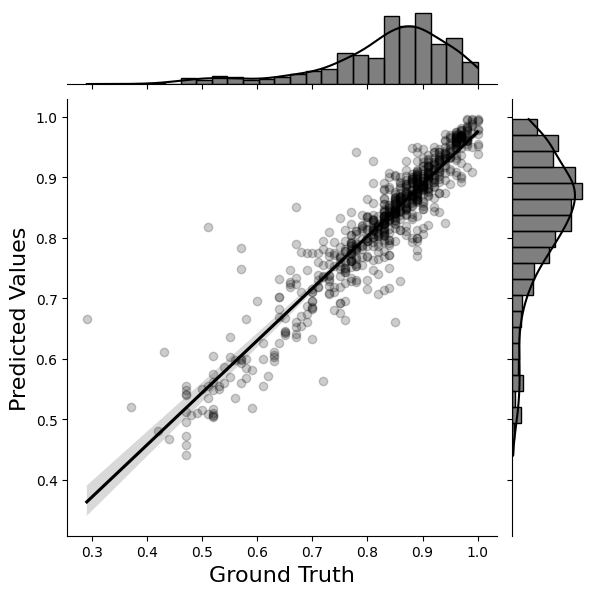}
  \caption{The R$^2$ of the model can be visually represented by the trend analysis of the ground truth data plotted against the predicted values. For each data point the values should be similar. Besides a few outliers, both ground truth and predicted distributions are similar.}
  \label{figR}
\end{figure}

The results of the ablation study are shown in table \ref{tab3}, which shows the R$^2$ of each model. We see that the SHAP method performed better than solely using feature importance from the LGBM model. Additionally, adding the ProtTrans embeddings also increases the model's R$^2$. The R$^2$ of the final model is shown in figure \ref{figR}. The sensitivity analysis done for each group of variables is also shown in section \ref{sensitivity} of the Supplementary Material.

\begin{table}[h!]
\centering
{%
\begin{tabular}{|l|l|l|}
\hline
\textbf{Model} & \textbf{Description} & \textbf{R²} \\ \hline
Model A & Baseline LGBM &   0.87730 \\ \hline
Model B & \begin{tabular}[c]{@{}l@{}}After LGBM feature importance selection\end{tabular} & 0.89256  \\ \hline
Model C & \begin{tabular}[c]{@{}l@{}}After LGBM feature importance selection\\ + transformer embeddings\end{tabular} &  0.90003 \\ \hline
Model D & After SHAP selection &   0.89901 \\ \hline
Model E & \begin{tabular}[c]{@{}l@{}}After SHAP selection \\ + transformer embeddings\end{tabular} &   0.90165 \\ \hline
\end{tabular}%
}
\caption{R$^2$ of the models trained on the ablation study.}
\label{tab3}
\end{table}





\pagebreak

\pagebreak
\section{Discussion}\label{sec4}
We propose an approach that allows us to predict the true ileal digestibility coefficient for each indispensable amino acid in a food, a value normally determined experimentally using \textit{in-vivo} or \textit{in-vitro} methods. Our data-science based method enabled us to estimate this value with 90\% R$^2$ accuracy. Not only is our proposed method novel in approximating the ileal digestibility coefficient of foods, as it is solely computational, but is also able to capture established relationships through its features and model output. Additionally, there are many potential downstream applications of our approach.
    
For instance, the approach itself can help accelerate the production of alternative protein products. Food scientists can use this approach to evaluate a large number of promising target proteins computationally and decrease the size of their experimental matrices. This methodology can also help predicting DIAAS for complex food matrices or for scenarios where \textit{in-vivo} and \textit{in-vitro} options cannot be used. In addition to the potential to streamline experimentation, our methodology's interpretability could incite scientists to propose new scientific experiments to better understand how protein structure is related to digestibility. We were able to capture some of these structure-activity relationships using the 20 most important features generated by Shapley value analysis. Our model's interpretability could guide \textit{in-vivo} experiments to better identify the direct relationship of these other variables with digestibility, reducing the array of possible experimental combinations when investigating these relationships. As new primary data is collected and becomes available, it is important perform regular re-trainings on our model to improve its explanatory power.

This highlights one of the fundamental challenges of our approach and the basis for our future research: the size of our curated dataset. As more data becomes available, we can retrain the model to capture less frequent patterns that are specific to local contexts, foods, and proteins. ML techniques often do not make assumptions on the exact functional form of the model and attempt to learn the model form directly from the data, such that it maximizes prediction accuracy. To do so, they take into account linear and non-linear correlations between the data, but do not take into account cause and effect relationships. In this situation, this means that the model learns patterns in the data based on the most commonly observed relationships across protein families. Therefore, with a  dataset with items ranging from processed foods to raw foods and meat analogues to animal products, more variability would be added and a generalizable model can be obtained. 

Much of the ground truth data used in this study was experimentally determined using the grown pig model, which is a valid animal model for this purpose \cite{moughan_piglet_1992, deglaire_animal_2012}. However, it has been established that food particle size impacts ileal digestibility values in pigs, with a decrease in particle size increasing apparent ileal digestibility \cite{flis_effect_2014}. This might introduce variance into the model, as the ileal digestibility values in our dataset were obtained from different experiments; there is likely limited consistency in food particle size across these experiments. 

Another avenue for future research is to add features related to food processing.
For instance, it is known that heat treatments applied to food can alter AA digestibility and bioavailability levels when compared to the same, untreated food \cite{hodgkinson_determination_2020,rutherfurd_application_1997,almeida_amino_2013,sa_food_2020}. Other values which we have not featurized are antinutritional factors, whose presence are known to decrease protein digestibility. For instance, tannins or polyphenols produce denaturation of proteins, and protease inhibitors can inhibit trypsin, pepsin, and other proteases in the gut, preventing digestion and absorption of proteins and amino acids. The dataset also fails to consider farming practices, which also affect the protein content of plants and livestock. For example, fertilization regimes can promote nutrients transfer to grains and changes in protein content \cite{preet_antinutrients_2000,yan_quantifying_2022}. Animal husbandry and nutrition also impact growth, productivity and protein content of animal products \cite{zhang_effect_2019,coulon_variations_1991,tyasi_assessing_2015}.

Overall, our model could be enriched with features that represent a food's structural information, antinutrients, processing, and the farming practices employed when it was growing. However, obtaining these data for a large number of food items is challenging. This demonstrates the strong need for a unified database that contains such information, more experiments to create ground truth values, and the need to iterate over our existing methodology to make it more robust once new data becomes available. In the meantime, using this purely computational, ethical, and affordable model can enable scientists to reduce the size of their experimental food matrices during protein digestibility experiments, especially in the development of new foods.

\section{Methods} \label{methods}

\subsection{Dataset creation details}

The final dataset obtained was created using the original 300 food items present on Muleya and Salter dataset \cite{molly_muleya_ileal_2021}. In cases where digestibility information was missing for some amino acids, an average value based on the reported amino acid data was assigned - average values are indicated in the dataset. 
Total amino acid and protein composition was obtained from the same source. Values were expressed in g/kg and recalculation was done in cases where the units reported in the reference were different. Digestible indispensable amino acids were calculated by multiplying the digestibility coefficient with the corresponding indispensable amino acid composition expressed in g/kg. Finally, digestible indispensable amino acids were expressed as mg/g protein by dividing the digestible indispensable amino acid composition with the total protein and converting to mg. DIAAS values were calculated for all age groups using the FAO recommended reference patterns. DIAAS values for some foods are missing due to insufficient information being reported in some cases. 

The nutritional information for foods was obtained from the FoodData Central of the United States Department of Agriculture (USDA) \cite{us_department_of_agriculture_usda_agricultural_research_service_nutrient_data_usda_nodate}. FoodData Central is an integrated data system that provides expanded nutrient profile data and links to related agricultural and experimental research. For each food item in the database, the features presented in tables \ref{table1} and \ref{table2} were added. 

In order to capture additional protein features, their physio-chemical and biochemical properties were added to the database.
For this, the amino acid FASTA sequence for each protein was first collected from UniProt \cite{bateman_uniprot_2021} or from the NCBI \cite{ncbi_national_nodate} protein database. For this, we selected up to three protein families of the food items, such as albumins, caseins and globulins, when present and Protlearn v2.1 \cite{dorfer_protlearn_nodate} was then used to extract the features from the FASTA sequences. This led to a total of 1671 features that were added to the database for each protein. All features obtained from Protlearn are present in section \ref{features_protlearn} of the Supplementary Material. 

Finally, the language model was used to obtain the embeddings of each protein family. The cleaned dataset was composed of almost 200 food items. 

\subsection{Baseline modeling approach}
First, baseline regressors on all 1671 features of 189 food items were trained. The first 3 top-performing models were based on boosting and bagging techniques of decision trees. These models are also able to deal with high-dimensional data that may contain non-linear effects and many interactions between covariates, and can be used to rank the most important predictors to gain insight into the resulting prediction model. Hyperparameters of these three models were optimized using a combination of a randomized grid search technique and manual tuning using stratified 5-fold cross-validation on the training set. Using the optimized tree-based models, the validation R$^{2}$ accuracy was calculated on a held-out test set. 

\subsection{SHAP}

Interpretability of models in AI is central to the practical impact of AI on society. Thus it can be treated as a problem of attribution. Shapley values \cite{kuhn_17_1953} provide the unique attribution method satisfying a set of intuitive axioms, e.g. they capture all interactions between features and sum to the model prediction. Given our baseline predictive model, SHAP yields a vector of importance scores associated with the underlying features. The Shapley value was first introduced as an axiomatic characterisation of a fair distribution of a total surplus from all players, and it may be used in predictive models where each feature is treated as a participant in the underlying game. While the Shapley value technique is conceptually appealing, it is also computationally demanding: in general, each Shapley value assessment demands an exponential number of model evaluations.

Classic Shapley values calculated through SHAP can be considered optimal since, within a large class of approaches, they are the only way to measure feature importance while maintaining several natural properties from cooperative game theory \cite{lundberg_local_2020}. Shapley also observes the value as an index for assessing the power of participants in a game, in addition to a model that seeks to predict the distribution of resources in multi-feature interactions. The value, like a price index or other market indices, aggregates the strength of actors in their numerous cooperation opportunities using averages (or weighted averages in certain extensions). The Shapley value may also be thought of as a measure of the utility of players/features in a game/model. Several studies have used Shapley values and SHAP package to improve the 
interpretability of tree-based models \cite{lundberg_local_2020,lundberg_explainable_2018}.

\subsection{ProtTrans}

The goal of language modelling is to estimate the probability distribution of various linguistic units, e.g., words, sentences. After training the language model (LM), we can extract some information learned by the LMs, referred to as embeddings. This process indicates that when a language model better understands the sequence, it also better understands the structure. 

Protein Language Models (pLMs) copy the concepts of Language Models from NLP by using tokens (words in NLP), i.e., amino acids from protein sequences, and treating entire proteins
like sentences in LMs. In step 1, these pLMs are trained in a self-supervised manner, essentially learning to predict masked amino acids (tokens) in already known sequences.

ProtTrans \cite{elnaggar_prottrans_2022} is a 3B-parameter model that has an encoder and decoder architecture and allows us to extract 1024 embeddings for each protein sequence using the 1B-parameter encoder. Thus, 1024 embeddings were generated for each of the three protein families of each food item.

\backmatter

\bibliographystyle{naturemag}
\bibliography{references_zotero_download}

\newpage
\section{Supplementary Material}

\subsection{Protein quality and bioavailability definition}

To define the pattern of human AA requirements, the FAO Committee on Protein Requirements \cite{fao_fao_1967} advised using a reference protein with an ideal AA composition.

The balance of AA across the small intestine (mouth to terminal ileum: ileal digestibility) or across the entire gut (mouth to anus: faecal digestibility) can provide a measure of the extent of digestion and absorption of food protein. Specifically, this balance is usually defined in terms of AA by the gastrointestinal tract for use by the body. Usually, the main interest of the analyses are the indispensable amino acids (IAAs), which cannot be synthesized by the human body and some dispensable amino acids (DAAs), such as arginine, cysteine, glutamine, glycine, proline and tyrosine, which can become conditionally indispensable for premature neonates \cite{efsa_scientific_2012}. 

Protein digestion is a long, complex and multi-stage process. The onset is from the mechanical process of breaking down food in the mouth and the chemical process of breaking down the molecular structure by salivary amylase. In the gastric phase of digestion, the resulting product is mixed with low pH gastric acid containing the protease pepsin. This allows the formation of (poly)peptides. In the small intestine, the chyme is mixed with pancreatic proteases and peptidases, such as trypsin, chymotrypsin, and carboxypeptidase A \cite{trommelen_gut_2021}. Together with intestinal brush border enzymes, these enzymes hydrolyse the proteins and (poly)peptides into amino acids, di-, tri-, and oligopeptides \cite{jahan-mihan_dietary_2011}. The amino acids and di- and tripeptides that are produced can be taken up by the small intestinal mucosa and are typically thought to be almost completely absorbed by the terminal ileum at the end of the small intestine \cite{van_der_wielen_presence_2021,adhikari_protein_2022}. It is important to mention that this is a unique process that depends on the individual gut microbiota, which plays and active role in digestion \cite{knight_gastrointestinal_2019}.

Peptides not absorbed in the ileum travel to the large intestine. However, there is no evidence that there is absorption of them after the ileum, and the absorbed values are small, not exceeding 0.1\% for whey protein in pigs \cite{van_der_wielen_presence_2021}. Many different methods for determining protein quality have been developed and utilized throughout the years. To quantify or characterize the protein quality of a protein supply, several approaches use distinct principles \cite{huang_review_2018}.

FAO \cite{shivakumar_protein_2020} suggested a ranking of different digesta based on representativeness to human digestibility for protein starts with human ileal digesta, followed by pig ileal digesta, rat ileal digesta, human feacal digesta, pig feacal digesta and rat feacal digesta.  Some human experiments have been conducted to test the ileal digestibility of proteins, however the  ileal digesta collection in animals is difficult \cite{gaudichon_determinants_2021}. Generally, pigs are used over humans because of the physiological similarity of their intestinal tract. Furthermore, there is a high correlation between ileal digesta from humans and pigs \cite{deglaire_animal_2012}. 

\begin{landscape}
\subsection{References of DIAAS experiments}

\begin{table}[h!]
\centering
\resizebox{\columnwidth}{!}{%
\begin{tabular}{|l|c|}
\hline
\multicolumn{1}{|c|}{\textbf{Paper}} &
  \textbf{Reference} \\ \hline
INRAE-CIRAD-AFZ Feed tables &
  \cite{feedtables} \\ \hline
The assessment of amino acid digestibility in foods for humans and including a collation of published ileal amino acid digestibility data for human foods &
  \cite{fao2011} \\ \hline
Ileal digestibility of intrinsically labelled hen's egg and meat protein determined with the dual stable isotope tracer method in Indian adults &
  \cite{kashyapr_2019} \\ \hline
Protein quality evaluation of complementary foods in Indian children &
  \cite{shivakumar_2019} \\ \hline
Cooking conditions affect the true ileal digestible amino acid content and digestible indispensable amino acid score (DIAAS) of bovine meat as determined in pigs &
  \cite{hodgkinson_2018} \\ \hline
Standardized ileal digestibility and digestible indispensable amino acid score of porcine and bovine hydrolysates in pigs &
  \cite{bindari_2018} \\ \hline
Digestible indispensable amino acid score and digestible amino acids in eight cereal grains &
  \cite{cervantes-pahm_digestible_2014} \\ \hline
True ilea amino acid digestibility and digestible indispensable amino acid scores (DIAASs) of plant-based protein foods &
  \cite{reynaud_true_2021} \\ \hline
Digestible indispensable amino acid scores of six cooked chinese pulses &
  \cite{han2020} \\ \hline
Very low ileal nitrogen and amino acid digestibility of zein compared to whey protein isolate in healthy volunteers &
  \cite{calvez2021} \\ \hline
Amino acid digestibility in housefly and black soldier fly prepupae by growing pigs &
  \cite{tan_2020} \\ \hline
Comparison of amino acid digestibility in full-fat soybean, two soybean meals and peanut flour between broiler chickens and growing pigs &
  \cite{park_2017} \\ \hline
Energy and amino acid digestibility of raw, steam-pelleted and extruded red lentil in growing pigs &
  \cite{hugman_2021} \\ \hline
Raw and roasted pistachio nuts are good sources of protein based on their digestible indispensable amino acid score as determined in pigs &
  \cite{bailey2020} \\ \hline
Effects of mealworm larvae hydrolysate on nutrient ileal digestibility in growing pigs compared to those of defatted mealworm larvae meal, fermented poultry byproduct and hydrolysed fish soluble &
  \cite{cho_effects_2020} \\ \hline
Evaluating standardized ileal digestibility of amino acids in growing pigs &
  \cite{yin_evaluating_2008} \\ \hline
True ileal digestibility of legumes determined by dual-isotope tracer method in Indian adults &
  \cite{kashyap_true_2019} \\ \hline
Amino acid digestibility of single cell protein from Corynebacterium &
  \cite{wang_amino_2013} \\ \hline
Chemical composition and nutritional values of feedstuffs &
  \cite{zziwa_2013} \\ \hline
Concentrations of digestible and metabolizable energy and amino acid digestibility by growing pigs may be reduced by autoclaving soybean meal &
  \cite{oliveira_concentrations_2020} \\ \hline
Values for DIAAS for some diary and plant proteins may better describe protein quality than values calculated using the concept for PDCAAS &
  \cite{mathai_values_2017} \\ \hline
Apparent and standardized true ileal digestibility of protein and amino acids from faba bean, lupin and pea, provided as whole seeds, dehulled or extruded in pig diets &
  \cite{mariscal-landin_apparent_2002} \\ \hline
Real ileal amino acid digestibility of pea protein compared to casein in healthy humans, a randomized trial &
  \cite{guillin_real_2022} \\ \hline
DIAAS is greater in animal based burgers than in plant based burgers if determined in pigs &
  \cite{fanelli2022} \\ \hline
The complementarity of amino acids in cooked pulse/cereal blends and effects on DIAAS &
  \cite{han_complementarity_2021} \\ \hline
Values for DIAAS determined in pigs are greater for milk than for breakfast cereals, but DIAAS values for individual ingredients are additive in combined meals &
  \cite{fanelli_values_2021} \\ \hline
Pinto bean amino acid digestibility and score in a Mexican dish with corn tortilla and guacamole, evaluated in adults using a dual tracer isotopic method &
  \cite{calderondelabarca_pinto_2021} \\ \hline
Most meat products have DIAAS that are greater than 100, but processing may increase or reduce protein quality &
  \cite{bailey_pork_2020} \\ \hline
Amino Acid Digestibility of Extruded Chickpea and Yellow Pea Protein is High and Comparable in Moderately Stunted South Indian Children with Use of a Dual Stable Isotope Tracer Method &
  \cite{devi_amino_2020} \\ \hline
Standardized ileal digestible amino acids and net energy contents in full fat and defatted black soldier fly larvae meals (Hermetia illucens) fed to growing pigs &
  \cite{crosbie_2020} \\ \hline
Digestibility of amino acids, fiber, and energy by growing pigs, and concentrations of digestible and metabolizable energy in yellow dent corn, hard red winter wheat, and sorghum may be influenced by extrusion &
  \cite{rodriguez_2020} \\ \hline
\end{tabular}%
}
\caption{References used to obtain the DIAAS values of almost 200 food items. }
\label{references}
\end{table}

\end{landscape}

\subsection{Features added to the dataset}

\begin{table}[h!]
\centering
\begin{tabular}{|c|c|}
\hline
\textbf{Feature}  & \textbf{Unit} \\ \hline
Protein           & g/kg          \\ \hline
Tryptophan (TRP)  & g/kg          \\ \hline
Threonine (THR)   & g/kg          \\ \hline
Isoleucine (ILE)  & g/kg          \\ \hline
Leucine (LEU)     & g/kg          \\ \hline
Lysine (LYS)      & g/kg          \\ \hline
Methionine (MET)  & g/kg          \\ \hline
Cysteine (CYS)    & g/kg          \\ \hline
Phenylaline (PHE) & g/kg          \\ \hline
Tyrosine (TYR)    & g/kg          \\ \hline
Valine (VAL)      & g/kg          \\ \hline
Arginine (ARG)    & g/kg          \\ \hline
Histidine (HIS)   & g/kg          \\ \hline
\end{tabular}
\vspace{0.3cm}
\caption{Total protein content and AAS are the features obtained for each food item in the training dataset.}
\label{table1}
\end{table}

\begin{table}[h!]
\centering
\begin{tabular}{|c|c|}
\hline
\textbf{Feature} & \textbf{Unit} \\ \hline
Energy           & kJ/100g       \\ \hline
Dietary Fiber    & g/100g        \\ \hline
Fat              & g/100g        \\ \hline
Ash              & g/100g        \\ \hline
Total Sugar      & g/100g        \\ \hline
Calcium          & mg/100g       \\ \hline
Sodium           & mg/100g       \\ \hline
Manganese        & mg/100g       \\ \hline
Zinc             & mg/100g        \\ \hline
Copper           & mg/100g       \\ \hline
Iron             & mg/100g       \\ \hline
Selenium         & ug/100g       \\ \hline
\end{tabular}
\vspace{0.3cm}
\caption{Nutritional information added to each food item in the training dataset.}
\label{table2}
\end{table}

\newpage

\pagebreak
\newpage
\subsection{Baseline models accuracy}\label{regressors}
Random forest is an ensemble of decision trees created by using bootstrap samples of the training dataset and random selection in tree induction \cite{breiman_random_2001}. XGBoost is an ensemble approach with a gradient descent–boosted decision tree algorithm \cite{chen_xgboost_2016}. LightGBM is an improvement framework based on the gradient descent–boosted decision tree algorithm and is more powerful than the previous XGBoost with a fast training speed and less memory occupation \cite{ke_lightgbm_2017}. 


\begin{table}[h!]
\centering
\begin{tabular}{|c|c|c|}
\hline
\textbf{Model}               & \textbf{R$^2$} & \textbf{RSME} \\ \hline
LGBM                         & 0.88           & 0.04          \\ \hline
XGBoost                      & 0.87           & 0.04          \\ \hline
Random Forest                & 0.82           & 0.05          \\ \hline
Gradient Boosting            & 0.82           & 0.05          \\ \hline
Bagging                      & 0.81           & 0.05          \\ \hline
KNeighbors                   & 0.73           & 0.06          \\ \hline
Decision Tree                & 0.72           & 0.06          \\ \hline
Nu Support Vector Regression & 0.70           & 0.07          \\ \hline
Linear Regression            & 0.66           & 0.07          \\ \hline
Lasso model (with Lars)      & 0.66           & 0.07          \\ \hline
SVR                          & 0.62           & 0.07          \\ \hline
Poisson                      & 0.49           & 0.09          \\ \hline
\end{tabular}%
\caption{Root mean squared error (RMSE) and R$^2$ for each baseline model. All the hyperparameters were set to the default parameters of Scikit-learn. No feature selection was applied i.e., all features were used.}

\end{table}

\pagebreak
\newpage
\subsection{Interpretability of feature impact for food item}

Using SHAP interaction values, we can decompose the impact 
of a feature on a specific sample, allowing for model interpretability. We selected three of the food items we predicted the true ileal digestibility and plot the contributions of each feature in the model for this specific sample, as shown in Supplementary figure \ref{shap_comparison}.

It is known that maize protein quality is fairly poor because it 
contains very low amounts of the essential amino threonine (THR) \cite{azevedo_biosynthesis_1997}. Tyrosine values were also not reported in our dataset, mainly because it's usually present in high amount on maize \cite{kohlmeier_nutrient_2003}. These and the dietary fiber content are negatively impacting the model. The hydrophobicity of the proteins of maize on the other hand, increase protein digestibility. It is known that an increase in protein hydrophobicity is directly related to protein unfolding and positively correlated with digestion \cite{he_effect_2018,tang_heat-induced_2022}. The same characteristic is observed on Pork. Thus, the small values of dietary fiber and magnesium of the food item also increased the protein digestibility. 

For potato protein concentrate, the amount of total protein is directly associated with the protein digestibility. The low values of dietary fiber also impacted the model positively and potatoes are usually high on lysine \cite{knorr_effect_1980}. Tofu's low levels of magnesium, potassium and dietary fiber increased the protein digestibility prediction. The original soy milk is heated during tofu's production process, which might lead to protein unfolding and an increase in protein hydrophobicity as well. 

\begin{landscape}
\begin{figure}[h!]
\centering
  \includegraphics[scale=1.2]{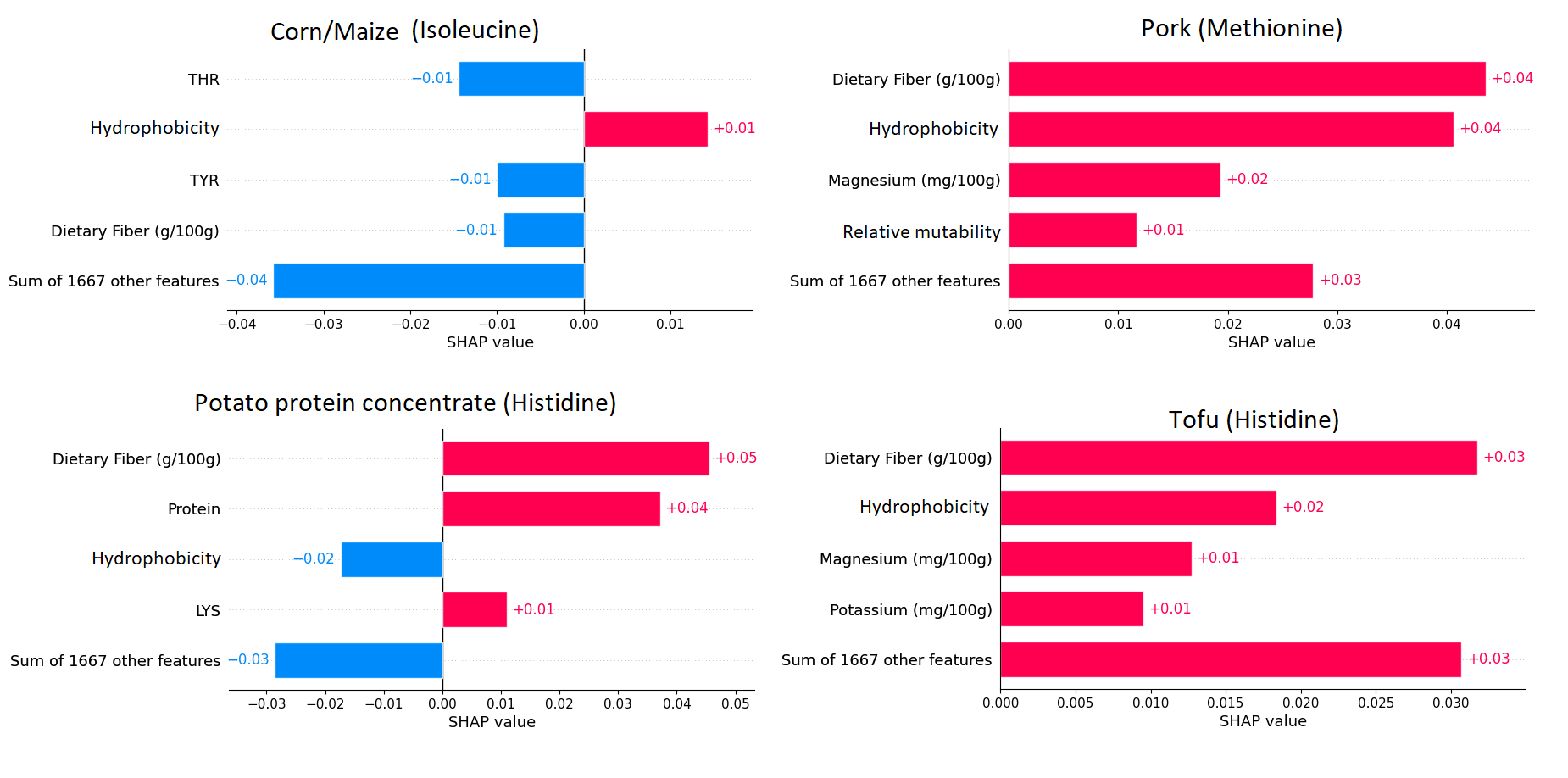}
  \caption{}
  \label{shap_comparison}
\end{figure}
\end{landscape}

\begin{landscape}
\subsection*{Limiting amino acid predicted for food items}

\begin{table}[h!]
\centering
\resizebox{\textwidth}{!}{%
\begin{tabular}{|c|ccc|}
\hline
\multicolumn{1}{|l|}{}      & \multicolumn{3}{c|}{\textbf{Limiting amino acid}}                                                                              \\ \hline
\textbf{Food Item}          & \multicolumn{1}{c|}{\textbf{DIAAS for Infants}} & \multicolumn{1}{c|}{\textbf{DIAAS for Children}} & \textbf{DIAAS for Adults} \\ \hline
Corn/Maize                  & \multicolumn{1}{c|}{Tryptophan}                 & \multicolumn{1}{c|}{Lysine}                      & Lysine                    \\ \hline
Soybean protein concentrate & \multicolumn{1}{c|}{Leucine}                    & \multicolumn{1}{c|}{Lysine}                      & Valine                    \\ \hline
Potato protein concentrate  & \multicolumn{1}{c|}{Tryptophan}                 & \multicolumn{1}{c|}{Histidine}                   & Histidine                 \\ \hline
Tofu                        & \multicolumn{1}{c|}{SAAs}                       & \multicolumn{1}{c|}{SAAs}                        & SAAs                      \\ \hline
Rye (Secale cereale L.)     & \multicolumn{1}{c|}{Tryptophan}                 & \multicolumn{1}{c|}{Lysine}                      & Lysine                    \\ \hline
Pork (raw belly)            & \multicolumn{1}{c|}{Tryptophan}                 & \multicolumn{1}{c|}{Valine}                      & Valine                    \\ \hline
Almond                      & \multicolumn{1}{c|}{Cysteine}                     & \multicolumn{1}{c|}{Cysteine}                      & Cysteine                   \\ \hline
\end{tabular}%
}
\caption{Limiting amino acid of each food item for each DIAAS category, considering the reference vales. SAAs represent the sulphur-containing amino acids.}
\label{tab7}
\end{table}

\end{landscape}
\pagebreak
\newpage
\subsection{Ablation study} \label{ablation}

\begin{landscape}
\begin{figure}[h!]
  \includegraphics[width=\linewidth]{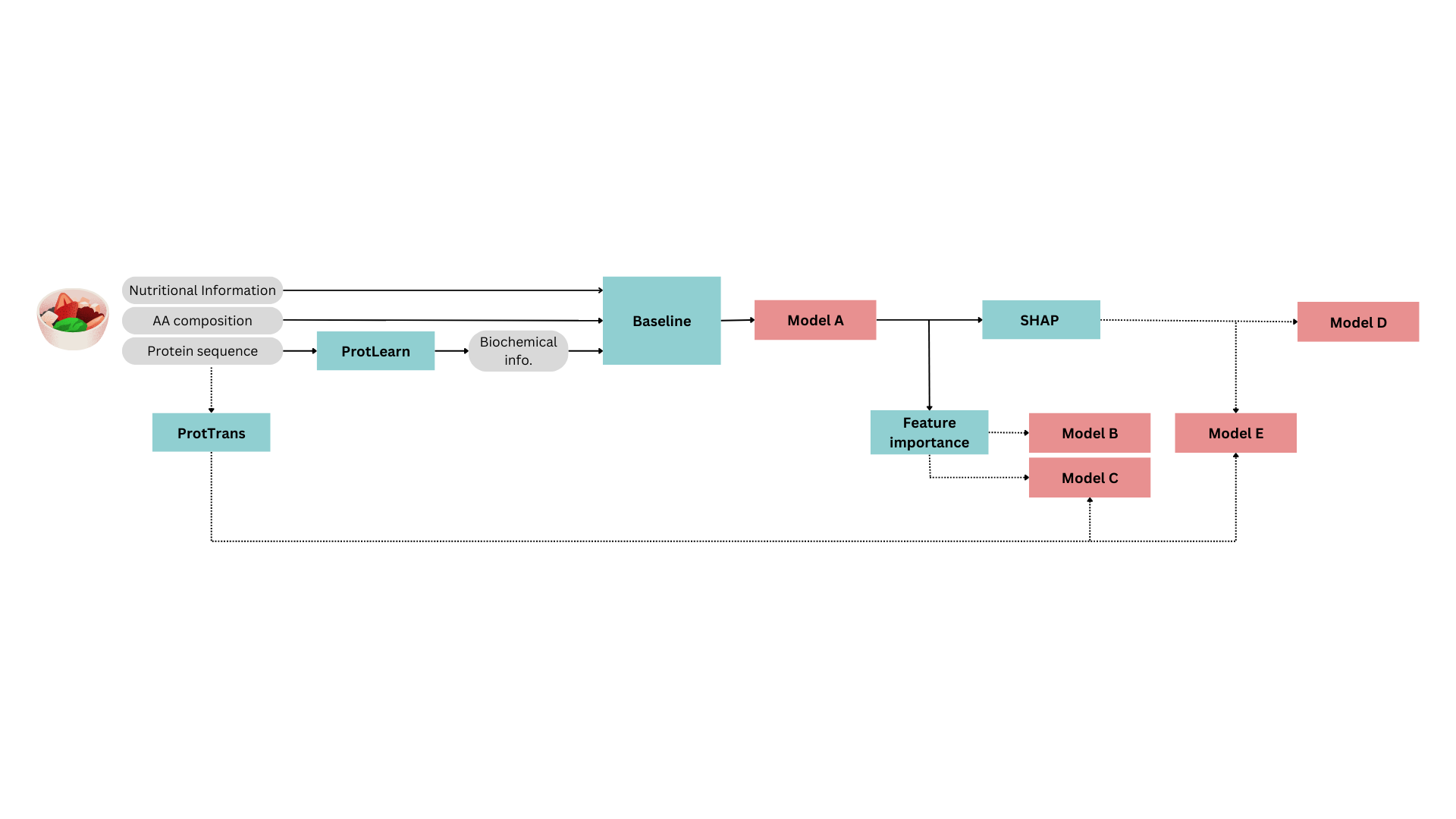}
  \caption{Architecture of the ablation study. The baseline model (A) is trained using nutritional information, AA composition and the biochemical features extracted from the protein sequences, using Protlearn. The most important features are selected using the feature importance from LGBM (model B) and the embeddings of the transformer model are added to the dataset (model C). Then, the feature importance is substituted by the Shapley (SHAP) method, when model D is trained without the embeddings and model E is trained with the embeddings.}
  \label{fig3}
\end{figure}
\end{landscape}

Figure \ref{fig3} shows the inputs that the different models were created with. For each model, we used 2304 samples, determined by the true ileal digestibility of the 12 indispensable aminoacids of 192 food items. Model A, also known as the baseline model, was trained using the complete dataset composed of 10700 features, such as nutritional information, amino acid score, and biochemical information obtained through protein sequence. Model B was trained with 25 features after the selection of the 20 most important features using SHAP. For model C we used these same 20 features with the 3 families of 1024 embeddings obtained from the language model, resulting in 3097 features. Hyperparameters of all these models were optimized using a combination of a randomized grid search technique and manual tuning using stratified 5-fold cross-validation on the training set.


\pagebreak
\newpage
\subsection{Sensitivity Analysis}\label{sensitivity}
\begin{figure}[h]
    \centering
    \includegraphics[width=\textwidth]{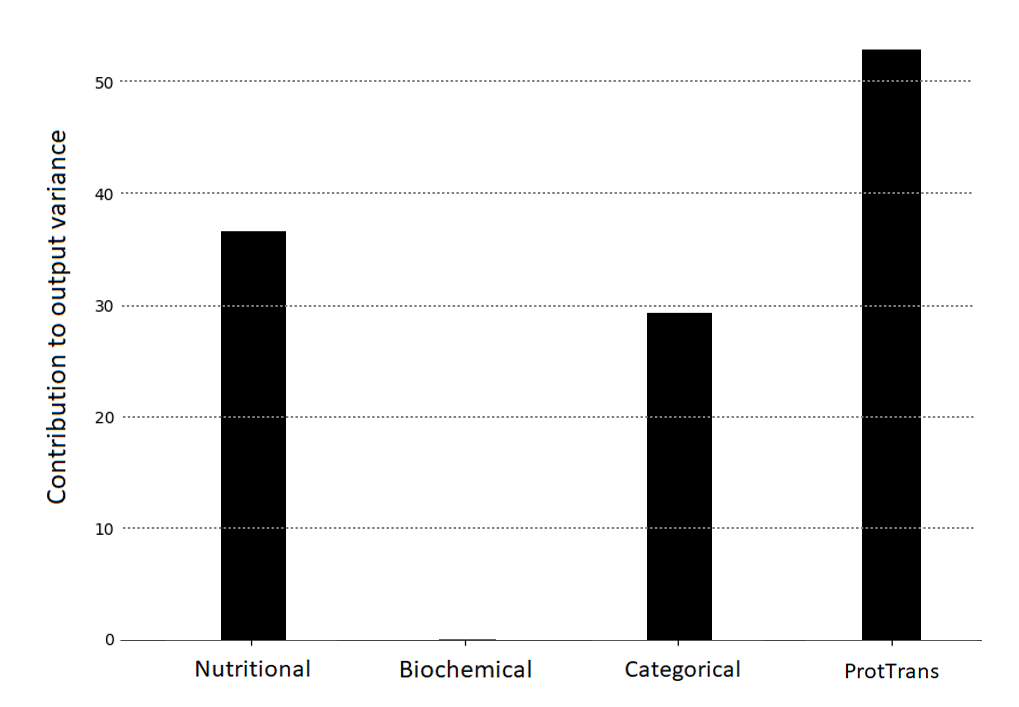}
    \caption{Results of sensitivity analysis. On the x axis are the inputs that we grouped as follows: nutritional, biochemical, categorical (including the protein families, food group and indispensable amino acid related features), and, ProTrans embeddings. The y axis shows the contribution of each of these input groups to the variance in the output i.e. the digestibility coefficient. }
    \label{fig:sen_analysis}
\end{figure}

In this section, we report the results of sensitivity analysis to study how the uncertainty in the model output is explained by the different input features. We use a variance based sensitivity analysis method as described in~\cite{SALTELLI2010259}. We used the SALib package~\cite{salib} for sensitivity analysis. Figure~\ref{fig:sen_analysis} illustrates the results of sensitivity analysis. We report the total order sensitivity effects that measure the variance in the output due to the first-order effects (by varying the input alone) and its interactions with other features. We observe that ProtTrans embeddings obtained by training on the protein sequences explained the maximum amount of variation (52\%) in the digestibility coefficient, our output. The remaining variation was contributed to by the nutritional (36\%), biochemical (0.04\%) and categorical features that include the protein families, food group and indispensable amino acid related inputs (29\%).

\pagebreak
\newpage
\subsection*{DIAAS value calculation}

To calculate the DIAAS value obtained through the predicted true ileal digestibility coefficient of each amino acid, the following equation is used:

\begin{equation}
    \text{DIAAS}[\%] = 100 \frac{\text{IAA}_T}{\text{IAA}_R}
\end{equation}

\noindent where IAA$_{T}$ is mg of digestible dietary IAA in 1 gram of the test protein and $IAA_{R}$ is the amount of mg of the same amino acid in 1 gram of the reference IAA. The digestible dietary content of each IAA is calculated as the content of each amino acid multiplied by their respective digestibility coefficient or true ileal digestibility score. The amount of each IAA in the reference protein is calculated by dividing the requirement for each IAA by the estimated average requirement. Thus, the DIAAS is determined by the most limiting IAA in the test protein in relation to its corresponding content in the reference protein. 

\begin{figure}[h!]
\centering
  \includegraphics[scale=1.3]{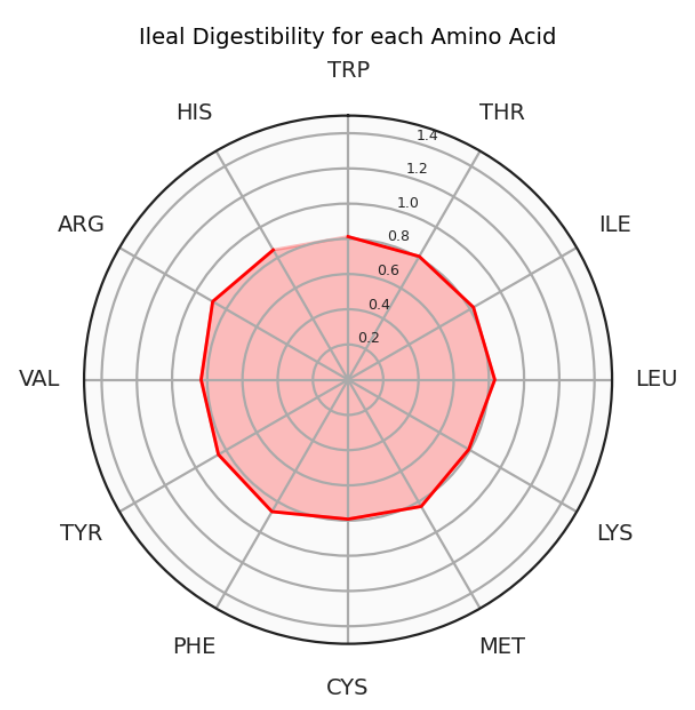}
  \caption{Radar plot of the true ileal digestibility for each amino acid for almond. The obtained values are 0.81 for tryptophan and threonine, 0.82 for isoleucine, 0.83 for leucine and methionine, 0.79 for lysine and cysteine, 0.87 for phenylalanine, 0.85 for valine and histidine and 0.89 for arginine.}
  \label{almond}
\end{figure}

Supplementary table \ref{tab6} shows the DIAAS for infants, children and adults calculated for 7 different food items. The ground truth values were obtained through experiments and referenced on the table. These values were calculated out of the predicted true ileal digestibility values for each amino acid. The limiting amino acids of each food item for each DIAAS category is shown in Supplementary table \ref{tab7}. An example of the calculated values for almond is shown in figure \ref{almond}.

To our knowledge, DIAAS has not been determined in almonds. However, PDCAAS values ranged from 44.3 to 47.8 \cite{house_determination_2019}. However, PDCAAS values were reported to be generally higher than a DIAAS values, especially for the poorer quality proteins. Therefore, having DIAAS values for as many food items as possible are of potential practical importance for populations in which dietary protein intake may be marginal \cite{rutherfurd_protein_2015}.

\begin{landscape}

\subsection*{DIAAS value of food items}
\begin{table}[h!]
\centering
{%
\begin{tabular}{|c|c|l|cc|cc|cc|}
\hline
\textbf{Food Item}          & \textbf{Origin} & \multicolumn{1}{c|}{ \textbf{Reference}} & \multicolumn{2}{c|}{\textbf{DIAAS for Infants}} & \multicolumn{2}{c|}{\textbf{DIAAS for Children}} & \multicolumn{2}{c|}{\textbf{DIAAS for Adults}} \\ \hline
                            &        & \multicolumn{1}{c|}{}          & Ground Truth                  & Predicted       & Ground Truth               & Predicted  & Ground Truth              & Predicted \\ \hline
Maize (yellow dent)                 & Train  &      \cite{cervantes-pahm_digestible_2014}                          & \multicolumn{1}{c|}{27}       & 29.91           & \multicolumn{1}{c|}{40}    & 39.65      & \multicolumn{1}{c|}{48}   & 47.09     \\ \hline
Soybean protein concentrate & Train  &     \cite{afz_amipig_2000}                           & \multicolumn{1}{c|}{76}       & 75.32           & \multicolumn{1}{c|}{102}   & 102.20     & \multicolumn{1}{c|}{112}  & 112.10    \\ \hline
Potato protein concentrate  & Test   &      \cite{afz_amipig_2000}                          & \multicolumn{1}{c|}{61}       & 68.26           & \multicolumn{1}{c|}{95}    & 93.73      & \multicolumn{1}{c|}{118}  & 117.17    \\ \hline
Tofu                        & Test   &      \cite{reynaud_true_2021}                          & \multicolumn{1}{c|}{62}       & 65.92           & \multicolumn{1}{c|}{75}    & 80.57      & \multicolumn{1}{c|}{88}   & 94.58     \\ \hline
Rye (Secale cereale L.)     & Test   &     \cite{cervantes-pahm_digestible_2014}                           & \multicolumn{1}{c|}{41}       & 40.69           & \multicolumn{1}{c|}{50}    & 51.85      & \multicolumn{1}{c|}{60}   & 61.57     \\ \hline
Pork (raw belly)            & Test   &          \cite{bailey_pork_2020}                      & \multicolumn{1}{c|}{69}       & 66.76           & \multicolumn{1}{c|}{110}   & 109.92     & \multicolumn{1}{c|}{119}  & 118.16    \\ \hline
Almond                      & -      &            -                    & \multicolumn{1}{l|}{No data available}         & 31.00           & \multicolumn{1}{c|}{No data available}      & 37.00      & \multicolumn{1}{c|}{No data available}     & 44.00     \\ \hline
\end{tabular}%
}
\caption{A few examples of DIAAS values predicted with our approach. Seven food items were selected from train, test and outside both train and test datasets. The ground truth for each was obtained from the references through experiments and the values of DIAAS for Infants, Children and Adults were calculated.}
\label{tab6}
\end{table}
\end{landscape}

\subsection{Biological and physiochemical features extracted from Protlearn}\label{features_protlearn}

The following table summarizes the features we extracted from the Protlearn package \cite{dorfer_protlearn_nodate}. For each food item in our  dataset, we collected 2 to 3 FASTA sequences of their most abundant protein families. By inputting each FASTA sequence into the Protlearn \cite{dorfer_protlearn_nodate} package, we obtained several features of that chemically and physically characterized the component protein families of each food. The large majority of these extracted features are Amino Acid Indices, a set of 566 indices  \cite{dorfer_protlearn_nodate} \cite{aaindex1} that represent various physiochemical properties of amino acids. The indices for each protein sequence are calculated by averaging the index values for all of the individual amino acids in the sequence.

\begin{table}[h!]
\begin{tabular}{|c|p{8cm}|}
\hline
    \textbf{Feature} & \textbf{Descriptor} \\ \hline
     Entropy & The Shannon entropy for the amino acid sequence \\ \hline
     C & Fraction of carbon atoms\\ \hline
     H & Fraction of hydrogen atoms \\ \hline
     N & Fraction of nitrogen atoms \\ \hline
     O & Fraction of oxygen atoms \\ \hline
     S & Fraction of sulfur atoms\\ \hline
     N & Fraction of nitrogen atoms\\ \hline
     Total Bound & Total number of bonds \\ \hline
     Single Bound & Number of single bonds \\ \hline
     Double Bound & Number of double bonds\\ \hline
     ANDN920101  & alpha-CH chemical shifts \\ \hline
    ARGP820101 & Hydrophobicity index \\ \hline
    ARGP820102 & Signal sequence helical potential \\ \hline
    ARGP820103 & Membrane-buried preference parameters \\ \hline
    BEGF750101 & Conformational parameter of inner helix \\ \hline
    BEGF750102 & Conformational parameter of beta-structure \\ \hline
    BEGF750103 & Conformational parameter of beta-turn \\ \hline
    BHAR880101 & Average flexibility indices \\ \hline
    BIGC670101 & Residue volume \\ \hline
    BIOV880101 & Information value for accessibility; average fraction 35\% \\ \hline
    BIOV880102 & Information value for accessibility; average fraction 23\% \\ \hline
    BROC820101 & Retention coefficient in TFA \\ \hline
    BROC820102 & Retention coefficient in HFBA \\ \hline
    BULH740101 & Transfer free energy to surface \\ \hline
    BULH740102 & Apparent partial specific volume \\ \hline
    BUNA790101 & alpha-NH chemical shifts \\ \hline
    BUNA790102 & alpha-CH chemical shifts \\ \hline
    BUNA790103 & Spin-spin coupling constants 3JHalpha-NH \\ \hline
    BURA740101 & Normalized frequency of alpha-helix \\ \hline
    BURA740102 & Normalized frequency of extended structure \\ \hline
    CHAM810101 & Steric parameter \\ \hline
    CHAM820101 & Polarizability parameter \\ \hline
    CHAM820102 & Free energy of solution in water, kcal/mole \\ \hline
    CHAM830101 & The Chou-Fasman parameter of the coil conformation \\ \hline
    CHAM830102 & A parameter defined from the residuals obtained from the best correlation of  the Chou-Fasman parameter of beta-sheet \\ \hline

\end{tabular}
\end{table}

\begin{table}[h!]
\begin{tabular}{|c|p{8cm}|}
\hline
    \textbf{Feature} & \textbf{Descriptor} \\ \hline
        CHAM830103 & The number of atoms in the side chain labelled $1+1$ \\ \hline
    CHAM830104 & The number of atoms in the side chain labelled $2+1$ \\ \hline
    CHAM830105 & The number of atoms in the side chain labelled $3+1$ \\ \hline
    CHAM830106 & The number of bonds in the longest chain \\ \hline
    CHAM830107 & A parameter of charge transfer capability \\ \hline
    CHAM830108 & A parameter of charge transfer donor capability \\ \hline
    CHOC750101 & Average volume of buried residue \\ \hline
    CHOC760101 & Residue accessible surface area in tripeptide \\ \hline
    CHOC760102 & Residue accessible surface area in folded protein \\ \hline
    CHOC760103 & Proportion of residues 95\% buried \\ \hline
    CHOC760104 & Proportion of residues 100\% buried \\ \hline
    CHOP780101 & Normalized frequency of beta-turn \\ \hline
    CHOP780201 & Normalized frequency of alpha-helix \\ \hline
    CHOP780202 & Normalized frequency of beta-sheet \\ \hline
    CHOP780203 & Normalized frequency of beta-turn \\ \hline
    CHOP780204 & Normalized frequency of N-terminal helix \\ \hline
    CHOP780205 & Normalized frequency of C-terminal helix \\ \hline
    CHOP780206 & Normalized frequency of N-terminal non helical region \\ \hline
    CHOP780207 & Normalized frequency of C-terminal non helical region \\ \hline
    CHOP780208 & Normalized frequency of N-terminal beta-sheet \\ \hline
    CHOP780209 & Normalized frequency of C-terminal beta-sheet \\ \hline
    CHOP780210 & Normalized frequency of N-terminal non beta region \\ \hline
    CHOP780211 & Normalized frequency of C-terminal non beta region \\ \hline
    CHOP780212 & Frequency of the 1st residue in turn \\ \hline
    CHOP780213 & Frequency of the 2nd residue in turn \\ \hline
    CHOP780214 & Frequency of the 3rd residue in turn \\ \hline
    CHOP780215 & Frequency of the 4th residue in turn \\ \hline
    CHOP780216 & Normalized frequency of the 2nd and 3rd residues in turn \\ \hline
    CIDH920101 & Normalized hydrophobicity scales for alpha-proteins \\ \hline
    CIDH920102 & Normalized hydrophobicity scales for beta-proteins \\ \hline
    CIDH920103 & Normalized hydrophobicity scales for alpha+beta-proteins \\ \hline
    CIDH920104 & Normalized hydrophobicity scales for alpha/beta-proteins \\ \hline
    CIDH920105 & Normalized average hydrophobicity scales \\ \hline
    COHE430101 & Partial specific volume \\ \hline

\end{tabular}
\end{table}

\begin{table}[h!]
\begin{tabular}{|c|p{8cm}|}
\hline
    \textbf{Feature} & \textbf{Descriptor} \\ \hline
       CIDH920104 & Normalized hydrophobicity scales for alpha/beta-proteins \\ \hline
    CIDH920105 & Normalized average hydrophobicity scales \\ \hline
    COHE430101 & Partial specific volume \\ \hline
    CRAJ730101 & Normalized frequency of middle helix \\ \hline
    CRAJ730102 & Normalized frequency of beta-sheet \\ \hline
    CRAJ730103 & Normalized frequency of turn \\ \hline
    DAWD720101 & Size \\ \hline
    DAYM780101 & Amino acid composition \\ \hline
    DAYM780201 & Relative mutability \\ \hline
    DESM900101 & Membrane preference for cytochrome b: MPH89 \\ \hline
    DESM900102 & Average membrane preference: AMP07 \\ \hline
    EISD840101 & Consensus normalized hydrophobicity scale \\ \hline
    EISD860101 & Solvation free energy \\ \hline
    EISD860102 & Atom-based hydrophobic moment \\ \hline
    EISD860103 & Direction of hydrophobic moment \\ \hline
    FASG760101 & Molecular weight \\ \hline
    FASG760102 & Melting point \\ \hline
    FASG760103 & Optical rotation \\ \hline
    FASG760104 & pK-N \\ \hline
    FASG760105 & pK-C \\ \hline
    FAUJ830101 & Hydrophobic parameter pi \\ \hline
    FAUJ880101 & Graph shape index \\ \hline
    FAUJ880102 & Smoothed upsilon steric parameter \\ \hline
    FAUJ880103 & Normalized van der Waals volume \\ \hline
    FAUJ880104 & STERIMOL length of the side chain \\ \hline
    FAUJ880105 & STERIMOL minimum width of the side chain \\ \hline
    FAUJ880106 & STERIMOL maximum width of the side chain \\ \hline
    FAUJ880107 & N.m.r. chemical shift of alpha-carbon \\ \hline
    FAUJ880108 & Localized electrical effect \\ \hline
    FAUJ880109 & Number of hydrogen bond donors \\ \hline
    FAUJ880110 & Number of full nonbonding orbitals \\ \hline
    FAUJ880111 & Positive charge \\ \hline
    FAUJ880112 & Negative charge \\ \hline
    FAUJ880113 & pK-a(RCOOH) \\ \hline
    FINA770101 & Helix-coil equilibrium constant \\ \hline
    FINA910101 & Helix initiation parameter at position $i-1$ \\ \hline
    FINA910102 & Helix initiation parameter at position $i$,$i+1$,$i+2$ \\ \hline
    FINA910103 & Helix termination parameter at position $j-2$,$j-1$,$j$ \\ \hline
    FINA910104 & Helix termination parameter at position $j+1$ \\ \hline
    GARJ730101 & Partition coefficient \\ \hline
    GEIM800101 & Alpha-helix indices \\ \hline

\end{tabular}
\end{table}

\begin{table}[h!]
\begin{tabular}{|c|p{8cm}|}
\hline
    \textbf{Feature} & \textbf{Descriptor} \\ \hline
    
        GEIM800102 & Alpha-helix indices for alpha-proteins \\ \hline
    GEIM800103 & Alpha-helix indices for beta-proteins \\ \hline
    GARJ730101 & Partition coefficient \\ \hline
    GEIM800101 & Alpha-helix indices \\ \hline
    GEIM800102 & Alpha-helix indices for alpha-proteins \\ \hline
    GEIM800103 & Alpha-helix indices for beta-proteins \\ \hline
    GEIM800104 & Alpha-helix indices for alpha/beta-proteins \\ \hline
    GEIM800105 & Beta-strand indices \\ \hline
    GEIM800106 & Beta-strand indices for beta-proteins \\ \hline
    GEIM800107 & Beta-strand indices for alpha/beta-proteins \\ \hline
    GEIM800108 & Aperiodic indices \\ \hline
    GEIM800109 & Aperiodic indices for alpha-proteins \\ \hline
    GEIM800110 & Aperiodic indices for beta-proteins \\ \hline
    GEIM800111 & Aperiodic indices for alpha/beta-proteins \\ \hline
    GOLD730101 & Hydrophobicity factor \\ \hline
    GOLD730102 & Residue volume \\ \hline
    GRAR740101 & Composition \\ \hline
    GRAR740102 & Polarity \\ \hline
    GRAR740103 & Volume \\ \hline
    GUYH850101 & Partition energy \\ \hline
    HOPA770101 & Hydration number \\ \hline
    HOPT810101 & Hydrophilicity value \\ \hline
    HUTJ700101 & Heat capacity \\ \hline
    HUTJ700102 & Absolute entropy \\ \hline
    HUTJ700103 & Entropy of formation \\ \hline
    ISOY800101 & Normalized relative frequency of alpha-helix \\ \hline
    ISOY800102 & Normalized relative frequency of extended structure \\ \hline
    ISOY800103 & Normalized relative frequency of bend \\ \hline
    ISOY800104 & Normalized relative frequency of bend R \\ \hline
    ISOY800105 & Normalized relative frequency of bend S \\ \hline
    ISOY800106 & Normalized relative frequency of helix end \\ \hline
    ISOY800107 & Normalized relative frequency of double bend \\ \hline
    ISOY800108 & Normalized relative frequency of coil \\ \hline
    JANJ780101 & Average accessible surface area \\ \hline
    JANJ780102 & Percentage of buried residues \\ \hline
    JANJ780103 & Percentage of exposed residues \\ \hline
    JANJ790101 & Ratio of buried and accessible molar fractions \\ \hline
    JANJ790102 & Transfer free energy \\ \hline
    JOND750101 & Hydrophobicity \\ \hline
JOND750102 & pK (-COOH) \\ \hline
    JOND920101 & Relative frequency of occurrence \\ \hline
    JOND920102 & Relative mutability \\ \hline

\end{tabular}
\end{table}

\begin{table}[h!]
\begin{tabular}{|c|p{8cm}|}
\hline
    \textbf{Feature} & \textbf{Descriptor} \\ \hline
       
    JUKT750101 & Amino acid distribution \\ \hline
    JUNJ780101 & Sequence frequency \\ \hline
    KANM800101 & Average relative probability of helix \\ \hline
    KANM800102 & Average relative probability of beta-sheet \\ \hline
    KANM800103 & Average relative probability of inner helix \\ \hline
    KANM800104 & Average relative probability of inner beta-sheet \\ \hline
    KARP850101 & Flexibility parameter for no rigid neighbors \\ \hline
    KARP850102 & Flexibility parameter for one rigid neighbor \\ \hline
    KARP850103 & Flexibility parameter for two rigid neighbors \\ \hline
    KHAG800101 & The Kerr-constant increments \\ \hline
    KLEP840101 & Net charge \\ \hline
    KRIW710101 & Side chain interaction parameter \\ \hline
    KRIW790101 & Side chain interaction parameter \\ \hline
    KRIW790102 & Fraction of site occupied by water \\ \hline
    KRIW790103 & Side chain volume \\ \hline
    KYTJ820101 & Hydropathy index \\ \hline
    LAWE840101 & Transfer free energy, CHP/water \\ \hline
    LEVM760101 & Hydrophobic parameter \\ \hline
    LEVM760102 & Distance between C-alpha and centroid of side chain \\ \hline
    LEVM760103 & Side chain angle theta(AAR) \\ \hline
    LEVM760104 & Side chain torsion angle phi(AAAR) \\ \hline
    LEVM760105 & Radius of gyration of side chain \\ \hline
    LEVM760106 & van der Waals parameter R0 \\ \hline
    LEVM760107 & van der Waals parameter epsilon \\ \hline
    LEVM780101 & Normalized frequency of alpha-helix, with weights \\ \hline
    LEVM780102 & Normalized frequency of beta-sheet, with weights \\ \hline
    LEVM780103 & Normalized frequency of reverse turn, with weights \\ \hline
    LEVM780104 & Normalized frequency of alpha-helix, unweighted \\ \hline
    LEVM780106 & Normalized frequency of reverse turn, unweighted \\ \hline

    LEWP710101 & Frequency of occurrence in beta-bends \\ \hline
    LIFS790101 & Conformational preference for all beta-strands\\ \hline
    LIFS790102 & Conformational preference for parallel beta-strands \\ \hline
    LIFS790103 & Conformational preference for antiparallel beta-strands \\ \hline
    MANP780101 & Average surrounding hydrophobicity \\ \hline
    MAXF760101 & Normalized frequency of alpha-helix \\ \hline
    MAXF760102 & Normalized frequency of extended structure \\ \hline
    MAXF760103 & Normalized frequency of zeta R \\ \hline
    MAXF760104 & Normalized frequency of left-handed alpha-helix\\ \hline
    MAXF760105 & Normalized frequency of zeta L \\ \hline
    MAXF760106 & Normalized frequency of alpha region\\ \hline
    MCMT640101 & Refractivity \\ \hline
    MEEJ800101 & Retention coefficient in HPLC, pH7.4 \\ \hline
    MEEJ800102 & Retention coefficient in HPLC, pH2.1 \\ \hline

\end{tabular}
\end{table}

\begin{table}[h!]
\begin{tabular}{|c|p{8cm}|}
\hline
    \textbf{Feature} & \textbf{Descriptor} \\ \hline
         MEEJ810101 & Retention coefficient in NaClO4 \\ \hline
    MEEJ810102 & Retention coefficient in NaH2PO4 \\ \hline
    MEIH800101 & Average reduced distance for C-alpha\\ \hline
    MEIH800102 & Average reduced distance for side chain \\ \hline
    MEIH800103 & Average side chain orientation angle \\ \hline
    MIYS850101 & Effective partition energy \\ \hline
    NAGK730101 & Normalized frequency of alpha-helix \\ \hline
    NAGK730102 & Normalized frequency of bata-structure \\ \hline
    NAGK730103 & Normalized frequency of coil \\ \hline
    NAKH900101 & AA composition of total proteins \\ \hline
    NAKH900102 & SD of AA composition of total proteins \\ \hline
    NAKH900103 & AA composition of mt-proteins \\ \hline
    NAKH900104 & Normalized composition of mt-proteins \\ \hline
    NAKH900105 & AA composition of mt-proteins from animal \\ \hline
    NAKH900106 & Normalized composition from animal \\ \hline
    NAKH900107 & AA composition of mt-proteins from fungi and plant \\ \hline
    NAKH900108 & Normalized composition from fungi and plant \\ \hline
    NAKH900109 & AA composition of membrane proteins \\ \hline
    NAKH900110 & Normalized composition of membrane proteins \\ \hline
    NAKH900111 & Transmembrane regions of non-mt-proteins \\ \hline
    NAKH900112 & Transmembrane regions of mt-proteins \\ \hline
    NAKH900113 & Ratio of average and computed composition \\ \hline
    NAKH920101 & AA composition of CYT of single-spanning proteins \\ \hline
    NAKH920102 & AA composition of CYT2 of single-spanning proteins \\ \hline
    NAKH920103 & AA composition of EXT of single-spanning proteins \\ \hline
    NAKH920104 & AA composition of EXT2 of single-spanning proteins \\ \hline
    NAKH920105 & AA composition of MEM of single-spanning proteins \\ \hline
    NAKH920106 & AA composition of CYT of multi-spanning proteins \\ \hline
    NAKH920107 & AA composition of EXT of multi-spanning proteins \\ \hline
    NAKH920108 & AA composition of MEM of multi-spanning proteins \\ \hline
    
    NISK800101 & 8 A contact number \\ \hline
    NISK860101 & 14 A contact number \\ \hline
    NOZY710101 & Transfer energy, organic solvent/water \\ \hline
    OOBM770101 & Average non-bonded energy per atom \\ \hline
    OOBM770102 & Short and medium range non-bonded energy per atom \\ \hline
    OOBM770103 & Long range non-bonded energy per atom \\ \hline
    OOBM770104 & Average non-bonded energy per residue \\ \hline
    OOBM770105 & Short and medium range non-bonded energy per residue \\ \hline
    OOBM850101 & Optimized beta-structure-coil equilibrium constant \\ \hline
    OOBM850102 & Optimized propensity to form reverse turn \\ \hline
    OOBM850103 & Optimized transfer energy parameter \\ \hline

\end{tabular}
\end{table}

\begin{table}[h!]

\begin{tabular}{|c|p{8cm}|}
\hline
    \textbf{Feature} & \textbf{Descriptor} \\ \hline
    OOBM850104 & Optimized average non-bonded energy per atom \\ \hline
    OOBM850105 & Optimized side chain interaction parameter \\ \hline
    PALJ810101 & Normalized frequency of alpha-helix from LG \\ \hline
    PALJ810102 & Normalized frequency of alpha-helix from CF \\ \hline
    PALJ810103 & Normalized frequency of beta-sheet from LG \\ \hline
    PALJ810104 & Normalized frequency of beta-sheet from CF \\ \hline
    PALJ810105 & Normalized frequency of turn from LG \\ \hline
    PALJ810106 & Normalized frequency of turn from CF \\ \hline
    PALJ810107 & Normalized frequency of alpha-helix in all-alpha class \\ \hline
    PALJ810108 & Normalized frequency of alpha-helix in alpha+beta class \\ \hline
    PALJ810109 & Normalized frequency of alpha-helix in alpha/beta class \\ \hline
    PALJ810110 & Normalized frequency of beta-sheet in all-beta class \\ \hline
    PALJ810111 & Normalized frequency of beta-sheet in alpha+beta class \\ \hline
    PALJ810112 & Normalized frequency of beta-sheet in alpha/beta class \\ \hline
    PALJ810113 & Normalized frequency of turn in all-alpha class \\ \hline
    PALJ810114 & Normalized frequency of turn in all-beta class \\ \hline
    PALJ810115 & Normalized frequency of turn in alpha+beta class \\ \hline
    PALJ810116 & Normalized frequency of turn in alpha/beta class \\ \hline
    PARJ860101 & HPLC parameter \\ \hline
    PLIV810101 & Partition coefficient \\ \hline
    PONP800101 & Surrounding hydrophobicity in folded form \\ \hline
    PONP800102 & Average gain in surrounding hydrophobicity \\ \hline
    PONP800103 & Average gain ratio in surrounding hydrophobicity \\ \hline
    PONP800104 & Surrounding hydrophobicity in alpha-helix \\ \hline
    PONP800105 & Surrounding hydrophobicity in beta-sheet \\ \hline
    PONP800106 & Surrounding hydrophobicity in turn \\ \hline
    PONP800107 & Accessibility reduction ratio \\ \hline
    PONP800108 & Average number of surrounding residues \\ \hline
    PRAM820101 & Intercept in regression analysis \\ \hline
    PRAM820102 & Slope in regression analysis x 1.0E1 \\ \hline

    PRAM820103 & Correlation coefficient in regression analysis \\ \hline
    PRAM900101 & Hydrophobicity \\ \hline
    PRAM900102 & Relative frequency in alpha-helix \\ \hline
    PRAM900103 & Relative frequency in beta-sheet \\ \hline
    PRAM900104 & Relative frequency in reverse-turn \\ \hline
    PTIO830101 & Helix-coil equilibrium constant \\ \hline

\end{tabular}
\end{table}

\begin{table}[h!]

\begin{tabular}{|c|p{8cm}|}
\hline
    \textbf{Feature} & \textbf{Descriptor} \\ \hline
  
    PONP800103 & Average gain ratio in surrounding hydrophobicity \\ \hline
    PONP800104 & Surrounding hydrophobicity in alpha-helix \\ \hline
    PONP800105 & Surrounding hydrophobicity in beta-sheet \\ \hline
    PONP800106 & Surrounding hydrophobicity in turn \\ \hline
    PONP800107 & Accessibility reduction ratio \\ \hline
    PONP800108 & Average number of surrounding residues \\ \hline
    PRAM820101 & Intercept in regression analysis \\ \hline
    PRAM820102 & Slope in regression analysis x 1.0E1 \\ \hline
    PTIO830102 & Beta-coil equilibrium constant \\ \hline
    QIAN880101 & Weights for alpha-helix at the window position of -6 \\ \hline
    QIAN880102 & Weights for alpha-helix at the window position of -5 \\ \hline
    QIAN880103 & Weights for alpha-helix at the window position of -4 \\ \hline
    QIAN880104 & Weights for alpha-helix at the window position of -3 \\ \hline
    QIAN880105 & Weights for alpha-helix at the window position of -2 \\ \hline
    QIAN880106 & Weights for alpha-helix at the window position of -1 \\ \hline
    QIAN880107 & Weights for alpha-helix at the window position of 0 \\ \hline
    QIAN880108 & Weights for alpha-helix at the window position of 1 \\ \hline
    QIAN880109 & Weights for alpha-helix at the window position of 2 \\ \hline
    QIAN880110 & Weights for alpha-helix at the window position of 3 \\ \hline
    QIAN880111 & Weights for alpha-helix at the window position of 4 \\ \hline
    QIAN880112 & Weights for alpha-helix at the window position of 5 \\ \hline
    QIAN880113 & Weights for alpha-helix at the window position of 6 \\ \hline
    QIAN880114 & Weights for beta-sheet at the window position of -6 \\ \hline
    QIAN880115 & Weights for beta-sheet at the window position of -5 \\ \hline
    QIAN880116 & Weights for beta-sheet at the window position of -4 \\ \hline
    QIAN880117 & Weights for beta-sheet at the window position of -3 \\ \hline
QIAN880118 & Weights for beta-sheet at the window position of -2 \\ \hline
    QIAN880119 & Weights for beta-sheet at the window position of -1 \\ \hline
    QIAN880120 & Weights for beta-sheet at the window position of 0 \\ \hline
    QIAN880121 & Weights for beta-sheet at the window position of 1 \\ \hline
    QIAN880122 & Weights for beta-sheet at the window position of 2 \\ \hline
    QIAN880123 & Weights for beta-sheet at the window position of 3 \\ \hline
    QIAN880124 & Weights for beta-sheet at the window position of 4 \\ \hline
    QIAN880125 & Weights for beta-sheet at the window position of 5 \\ \hline
    QIAN880126 & Weights for beta-sheet at the window position of 6 \\ \hline
    QIAN880127 & Weights for coil at the window position of -6 \\ \hline
    QIAN880128 & Weights for coil at the window position of -5 \\ \hline
    QIAN880129 & Weights for coil at the window position of -4 \\ \hline
    QIAN880130 & Weights for coil at the window position of -3 \\ \hline
    QIAN880131 & Weights for coil at the window position of -2 \\ \hline
    QIAN880132 & Weights for coil at the window position of -1 \\ \hline
    QIAN880133 & Weights for coil at the window position of 0 \\ \hline

\end{tabular}
\end{table}

\begin{table}[h!]

\begin{tabular}{|c|p{8cm}|}
\hline
    \textbf{Feature} & \textbf{Descriptor} \\ \hline
  
    QIAN880134 & Weights for coil at the window position of 1 \\ \hline
    QIAN880135 & Weights for coil at the window position of 2 \\ \hline
    QIAN880136 & Weights for coil at the window position of 3 \\ \hline
    QIAN880137 & Weights for coil at the window position of 4 \\ \hline
    QIAN880138 & Weights for coil at the window position of 5 \\ \hline
    QIAN880139 & Weights for coil at the window position of 6 \\ \hline
    RACS770101 & Average reduced distance for C-alpha \\ \hline
    RACS770102 & Average reduced distance for side chain \\ \hline
    RACS770103 & Side chain orientational preference \\ \hline
    RACS820101 & Average relative fractional occurrence in A0(i) \\ \hline
    RACS820102 & Average relative fractional occurrence in AR(i) \\ \hline
    RACS820103 & Average relative fractional occurrence in AL(i) \\ \hline
    RACS820104 & Average relative fractional occurrence in EL(i) \\ \hline
    RACS820105 & Average relative fractional occurrence in E0(i) \\ \hline
    RACS820106 & Average relative fractional occurrence in ER(i) \\ \hline
    RACS820107 & Average relative fractional occurrence in A0(i-1) \\ \hline
    RACS820108 & Average relative fractional occurrence in AR(i-1) \\ \hline
    RACS820109 & Average relative fractional occurrence in AL(i-1) \\ \hline
    RACS820110 & Average relative fractional occurrence in EL(i-1) \\ \hline
    RACS820111 & Average relative fractional occurrence in E0(i-1) \\ \hline
    RACS820112 & Average relative fractional occurrence in ER(i-1) \\ \hline
    RACS820113 & Value of theta(i) \\ \hline
    RACS820114 & Value of theta(i-1) \\ \hline
    RADA880101 & Transfer free energy from chx to wat \\ \hline
    RADA880102 & Transfer free energy from oct to wat \\ \hline
    RADA880103 & Transfer free energy from vap to chx \\ \hline
    RADA880104 & Transfer free energy from chx to oct \\ \hline
    RADA880105 & Transfer free energy from vap to oct \\ \hline
    RADA880106 & Accessible surface area \\ \hline
    RADA880107 & Energy transfer from out to in(95\%buried) \\ \hline
    RADA880108 & Mean polarity \\ \hline
    RICJ880101 & Relative preference value at N" \\ \hline
    RICJ880102 & Relative preference value at N' \\ \hline
    RICJ880103 & Relative preference value at N-cap \\ \hline
    RICJ880104 & Relative preference value at N1 \\ \hline
    RICJ880105 & Relative preference value at N2 \\ \hline
RICJ880106 & Relative preference value at N3 \\ \hline
    RICJ880107 & Relative preference value at N4 \\ \hline
    RICJ880108 & Relative preference value at N5 \\ \hline
    RICJ880109 & Relative preference value at Mid \\ \hline
    RICJ880110 & Relative preference value at C5 \\ \hline
    RICJ880111 & Relative preference value at C4 \\ \hline
    RICJ880112 & Relative preference value at C3 \\ \hline

\end{tabular}
\end{table}

\begin{table}[h!]
\begin{tabular}{|c|p{8cm}|}
\hline
    \textbf{Feature} & \textbf{Descriptor} \\ \hline
  RICJ880113 & Relative preference value at C2 \\ \hline
    RICJ880114 & Relative preference value at C1 \\ \hline
    RICJ880115 & Relative preference value at C-cap \\ \hline
    RICJ880116 & Relative preference value at C' \\ \hline
    RICJ880117 & Relative preference value at C"\\ \hline
    ROBB760101 & Information measure for alpha-helix \\ \hline
    ROBB760102 & Information measure for N-terminal helix \\ \hline
    ROBB760103 & Information measure for middle helix \\ \hline
    ROBB760104 & Information measure for C-terminal helix \\ \hline
    ROBB760105 & Information measure for extended \\ \hline
    ROBB760106 & Information measure for pleated-sheet \\ \hline
    ROBB760107 & Information measure for extended without H-bond \\ \hline
    ROBB760108 & Information measure for turn \\ \hline
    ROBB760109 & Information measure for N-terminal turn \\ \hline
    ROBB760110 & Information measure for middle turn \\ \hline
    ROBB760111 & Information measure for C-terminal turn \\ \hline
    ROBB760112 & Information measure for coil \\ \hline
    ROBB760113 & Information measure for loop \\ \hline
    ROBB790101 & Hydration free energy\\ \hline
    ROSG850101 & Mean area buried on transfer \\ \hline
    ROSG850102 & Mean fractional area loss \\ \hline
    ROSM880101 & Side chain hydropathy, uncorrected for solvation \\ \hline
    ROSM880102 & Side chain hydropathy, corrected for solvation \\ \hline
    ROSM880103 & Loss of Side chain hydropathy by helix formation \\ \hline
    SIMZ760101 & Transfer free energy \\ \hline
    SNEP660101 & Principal component I \\ \hline
    SNEP660102 & Principal component II \\ \hline
    SNEP660103 & Principal component III \\ \hline
    SNEP660104 & Principal component IV \\ \hline
    SUEM840101 & Zimm-Bragg parameter s at 20 C \\ \hline
    SUEM840102 & Zimm-Bragg parameter sigma x 1.0E4 \\ \hline
    SWER830101 & Optimal matching hydrophobicity \\ \hline
    TANS770101 & Normalized frequency of alpha-helix \\ \hline
    TANS770102 & Normalized frequency of isolated helix \\ \hline
    TANS770103 & Normalized frequency of extended structure \\ \hline
    TANS770104 & Normalized frequency of chain reversal R \\ \hline
    TANS770105 & Normalized frequency of chain reversal S \\ \hline
    TANS770106 & Normalized frequency of chain reversal D \\ \hline
    TANS770107 & Normalized frequency of left-handed helix \\ \hline
    TANS770108 & Normalized frequency of zeta R \\ \hline
    TANS770109 & Normalized frequency of coil \\ \hline
    TANS770110 & Normalized frequency of chain reversal \\ \hline

\end{tabular}
\end{table}

\begin{table}[h!]

\begin{tabular}{|c|p{8cm}|}
\hline
    \textbf{Feature} & \textbf{Descriptor} \\ \hline
  VASM830101 & Relative population of conformational state A \\ \hline
    VASM830102 & Relative population of conformational state C \\ \hline
    VASM830103 & Relative population of conformational state E \\ \hline
    VELV850101 & Electron-ion interaction potential \\ \hline
    VENT840101 & Bitterness \\ \hline
    VHEG790101 & Transfer free energy to lipophilic phase\\ \hline
    WARP780101 & Average interactions per side chain atom \\ \hline
    WEBA780101 & RF value in high salt chromatography\\ \hline
    WERD780101 & Propensity to be buried inside \\ \hline
    WERD780102 & Free energy change of epsilon(i) to epsilon(ex) \\ \hline
    WERD780103 & Free energy change of alpha(Ri) to alpha(Rh) \\ \hline
    WERD780104 & Free energy change of epsilon(i) to alpha(Rh) \\ \hline
    WOEC730101 & Polar requirement\\ \hline
    WOLR810101 & Hydration potential\\ \hline
    WOLS870101 & Principal property value z1 \\ \hline
    WOLS870102 & Principal property value z2 \\ \hline
    WOLS870103 & Principal property value z3 \\ \hline
    YUTK870101 & Unfolding Gibbs energy in water, pH7.0 \\ \hline
    YUTK870102 & Unfolding Gibbs energy in water, pH9.0 \\ \hline
    YUTK870103 & Activation Gibbs energy of unfolding, pH7.0 \\ \hline
    YUTK870104 & Activation Gibbs energy of unfolding, pH9.0 \\ \hline
    ZASB820101 & Dependence of partition coefficient on ionic strength \\ \hline
    ZIMJ680101 & Hydrophobicity \\ \hline
    ZIMJ680102 & Bulkiness \\ \hline
    ZIMJ680103 & Polarity \\ \hline
    ZIMJ680104 & Isoelectric point \\ \hline
    ZIMJ680105 & RF rank \\ \hline
    AURR980101 & Normalized positional residue frequency at helix termini N4'\\ \hline
    AURR980102 & Normalized positional residue frequency at helix termini N"' \\ \hline
    AURR980103 & Normalized positional residue frequency at helix termini N" \\ \hline
    AURR980104 & Normalized positional residue frequency at helix termini N' \\ \hline
    AURR980105 & Normalized positional residue frequency at helix termini Nc \\ \hline
    AURR980106 & Normalized positional residue frequency at helix termini N1 \\ \hline
    AURR980107 & Normalized positional residue frequency at helix termini N2 \\ \hline
    AURR980108 & Normalized positional residue frequency at helix termini N3 \\ \hline
    AURR980109 & Normalized positional residue frequency at helix termini N4 \\ \hline

\end{tabular}
\end{table}

\begin{table}[h!]

\begin{tabular}{|c|p{8cm}|}
\hline
    \textbf{Feature} & \textbf{Descriptor} \\ \hline
  
         AURR980110 & Normalized positional residue frequency at helix termini N5 \\ \hline
    AURR980111 & Normalized positional residue frequency at helix termini C5 \\ \hline
    AURR980112 & Normalized positional residue frequency at helix termini C4 \\ \hline  
    
    AURR980113 & Normalized positional residue frequency at helix termini C3 \\ \hline
    AURR980114 & Normalized positional residue frequency at helix termini C2 \\ \hline
    AURR980115 & Normalized positional residue frequency at helix termini C1 \\ \hline
    AURR980116 & Normalized positional residue frequency at helix termini Cc \\ \hline
    AURR980117 & Normalized positional residue frequency at helix termini C' \\ \hline
    AURR980118 & Normalized positional residue frequency at helix termini C" \\ \hline
    AURR980119 & Normalized positional residue frequency at helix termini C"' \\ \hline
    AURR980120 & Normalized positional residue frequency at helix termini C4' \\ \hline
    ONEK900101 & Delta G values for the peptides extrapolated to 0 M urea \\ \hline
    ONEK900102 & Helix formation parameters (delta delta G) \\ \hline
    VINM940101 & Normalized flexibility parameters (B-values), average \\ \hline
    VINM940102 & Normalized flexibility parameters (B-values) for each residue surrounded by  none rigid neighbours \\ \hline
    VINM940103 & Normalized flexibility parameters (B-values) for each residue surrounded by  one rigid neighbours \\ \hline
    VINM940104 & Normalized flexibility parameters (B-values) for each residue surrounded by  two rigid neighbours \\ \hline
    MUNV940101 & Free energy in alpha-helical conformation \\ \hline
    MUNV940102 & Free energy in alpha-helical region \\ \hline
    MUNV940103 & Free energy in beta-strand conformation \\ \hline
    MUNV940104 & Free energy in beta-strand region \\ \hline
    MUNV940105 & Free energy in beta-strand region \\ \hline
    WIMW960101 & Free energies of transfer of AcWl-X-LL peptides from bilayer interface to  water \\ \hline
    KIMC930101 & Thermodynamic beta sheet propensity \\ \hline
    MONM990101 & Turn propensity scale for transmembrane helices \\ \hline
    BLAM930101 & Alpha helix propensity of position 44 in T4 lysozyme \\ \hline

\end{tabular}
\end{table}

\begin{table}[h!]
\begin{tabular}{|c|p{8cm}|}
\hline
    \textbf{Feature} & \textbf{Descriptor} \\ \hline
      PARS000101 & p-Values of mesophilic proteins based on the distributions of B values  \\ \hline
    PARS000102 & p-Values of thermophilic proteins based on the distributions of B values  \\ \hline

        KUMS000101 & Distribution of amino acid residues in the 18 non-redundant families of  thermophilic proteins \\ \hline
    KUMS000102 & Distribution of amino acid residues in the 18 non-redundant families of  mesophilic proteins \\ \hline
    KUMS000103 & Distribution of amino acid residues in the alpha-helices in thermophilic  proteins \\ \hline
    KUMS000104 & Distribution of amino acid residues in the alpha-helices in mesophilic  proteins \\ \hline
    TAKK010101 & Side-chain contribution to protein stability (kJ/mol) \\ \hline
    FODM020101 & Propensity of amino acids within pi-helices \\ \hline
    NADH010101 & Hydropathy scale based on self-information values in the two-state model (5\%  accessibility) \\ \hline
    NADH010102 & Hydropathy scale based on self-information values in the two-state model (9\%  accessibility) \\ \hline
    NADH010103 & Hydropathy scale based on self-information values in the two-state model (16\%  accessibility) \\ \hline
    NADH010104 & Hydropathy scale based on self-information values in the two-state model (20\%  accessibility) \\ \hline
    NADH010105 & Hydropathy scale based on self-information values in the two-state model (25\%  accessibility) \\ \hline
    NADH010106 & Hydropathy scale based on self-information values in the two-state model (36\%  accessibility) \\ \hline
    NADH010107 & Hydropathy scale based on self-information values in the two-state model (50\%  accessibility) \\ \hline
    MONM990201 & Averaged turn propensities in a transmembrane helix\\ \hline
    KOEP990101 & Alpha-helix propensity derived from designed sequences \\ \hline
    KOEP990102 & Beta-sheet propensity derived from designed sequences \\ \hline
    CEDJ970101 & Composition of amino acids in extracellular proteins (percent) \\ \hline
    CEDJ970102 & Composition of amino acids in anchored proteins (percent) \\ \hline
    CEDJ970103 & Composition of amino acids in membrane proteins (percent) \\ \hline
    CEDJ970104 & Composition of amino acids in intracellular proteins (percent) \\ \hline
    CEDJ970105 & Composition of amino acids in nuclear proteins (percent) \\ \hline

\end{tabular}
\end{table}

  \begin{table}[h!]
\begin{tabular}{|c|p{8cm}|}
\hline
    \textbf{Feature} & \textbf{Descriptor} \\ \hline
  FUKS010101 & Surface composition of amino acids in intracellular proteins of thermophiles  (percent) \\ \hline
    FUKS010102 & Surface composition of amino acids in intracellular proteins of mesophiles  (percent) \\ \hline
    FUKS010103 & Surface composition of amino acids in extracellular proteins of mesophiles  (percent) \\ \hline
    FUKS010104 & Surface composition of amino acids in nuclear proteins (percent)  \\ \hline
    FUKS010105 & Interior composition of amino acids in intracellular proteins of thermophiles  (percent) \\ \hline
    FUKS010106 & Interior composition of amino acids in intracellular proteins of mesophiles  (percent) \\ \hline
    FUKS010107 & Interior composition of amino acids in extracellular proteins of mesophiles  (percent) \\ \hline
    FUKS010108 & Interior composition of amino acids in nuclear proteins (percent)  \\ \hline
    FUKS010109 & Entire chain composition of amino acids in intracellular proteins of  thermophiles (percent) \\ \hline
    FUKS010110 & Entire chain composition of amino acids in intracellular proteins of  mesophiles (percent) \\ \hline
    FUKS010111 & Entire chain composition of amino acids in extracellular proteins of  mesophiles (percent) \\ \hline
    FUKS010112 & Entire chain compositino of amino acids in nuclear proteins (percent) \\ \hline
    MITS020101 & Amphiphilicity index \\ \hline
    TSAJ990101 & Volumes including the crystallographic waters using the ProtOr \\ \hline
    TSAJ990102 & Volumes not including the crystallographic waters using the ProtOr \\ \hline
    COSI940101 & Electron-ion interaction potential values \\ \hline
    PONP930101 & Hydrophobicity scales \\ \hline
    WILM950101 & Hydrophobicity coefficient in RP-HPLC, C18 with 0.1\%TFA/MeCN/H$_2$O \\ \hline
    WILM950102 & Hydrophobicity coefficient in RP-HPLC, C8 with 0.1\%TFA/MeCN/H$_2$O \\ \hline
    WILM950103 & Hydrophobicity coefficient in RP-HPLC, C4 with 0.1\%TFA/MeCN/H$_2$O \\ \hline
    WILM950104 & Hydrophobicity coefficient in RP-HPLC, C18 with 0.1\%TFA/2-PrOH/MeCN/H$_2$O  \\ \hline
        KUHL950101 & Hydrophilicity scale \\ \hline
    GUOD860101 & Retention coefficient at pH 2\\ \hline
    JURD980101 & Modified Kyte-Doolittle hydrophobicity scale \\ \hline

\end{tabular}
\end{table}

    \begin{table}[h!]
\begin{tabular}{|c|p{8cm}|}
\hline
    \textbf{Feature} & \textbf{Descriptor} \\ \hline
  BASU050101 & Interactivity scale obtained from the contact matrix \\ \hline
    BASU050102 & Interactivity scale obtained by maximizing the mean of correlation  coefficient over single-domain globular proteins \\ \hline
    BASU050103 & Interactivity scale obtained by maximizing the mean of correlation  coefficient over pairs of sequences sharing the TIM barrel fold \\ \hline
    SUYM030101 & Linker propensity index \\ \hline
    PUNT030101 & Knowledge-based membrane-propensity scale from 1D Helix in MPtopo databases  \\ \hline
    PUNT030102 & Knowledge-based membrane-propensity scale from 3D Helix in MPtopo databases  \\ \hline
    GEOR030101 & Linker propensity from all dataset \\ \hline
    GEOR030102 & Linker propensity from 1-linker dataset \\ \hline
    GEOR030103 & Linker propensity from 2-linker dataset \\ \hline
    GEOR030104 & Linker propensity from 3-linker dataset \\ \hline
    GEOR030105 & Linker propensity from small dataset (linker length is less than six  residues) \\ \hline
    GEOR030106 & Linker propensity from medium dataset (linker length is between six and 14  residues) \\ \hline
        GEOR030107 & Linker propensity from long dataset (linker length is greater than 14  residues) \\ \hline
    GEOR030108 & Linker propensity from helical (annotated by DSSP) dataset \\ \hline
    GEOR030109 & Linker propensity from non-helical (annotated by DSSP) dataset  \\ \hline
    ZHOH040101 & The stability scale from the knowledge-based atom-atom potential \\ \hline
    ZHOH040102 & The relative stability scale extracted from mutation experiments \\ \hline
    ZHOH040103 & Buriability \\ \hline
    BAEK050101 & Linker index \\ \hline
    HARY940101 & Mean volumes of residues buried in protein interiors \\ \hline
    PONJ960101 & Average volumes of residues \\ \hline
DIGM050101 & Hydrostatic pressure asymmetry index, PAI \\ \hline
    WOLR790101 & Hydrophobicity index \\ \hline
    OLSK800101 & Average internal preferences \\ \hline
    KIDA850101 & Hydrophobicity-related index \\ \hline
    GUYH850102 & Apparent partition energies calculated from Wertz-Scheraga index\\ \hline
    GUYH850103 & Apparent partition energies calculated from Robson-Osguthorpe index \\ \hline)
    GUYH850104 & Apparent partition energies calculated from Janin index \\ \hline

\end{tabular}
\end{table}

\begin{table}[h!]
\begin{tabular}{|c|p{8cm}|}
\hline
    \textbf{Feature} & \textbf{Descriptor} \\ \hline
 GUYH850105 & Apparent partition energies calculated from Chothia index \\ \hline
    JACR890101 & Weights from the IFH scale \\ \hline
    COWR900101 & Hydrophobicity index, 3.0 pH \\ \hline
    BLAS910101 & Scaled side chain hydrophobicity values \\ \hline
    CASG920101 & Hydrophobicity scale from native protein structures \\ \hline
    CORJ870101 & NNEIG index \\ \hline
    CORJ870102 & SWEIG index \\ \hline
    CORJ870103 & PRIFT index \\ \hline
    CORJ870104 & PRILS index \\ \hline
    CORJ870105 & ALTFT index \\ \hline
    CORJ870106 & ALTLS index \\ \hline
    CORJ870107 & TOTFT index \\ \hline
    CORJ870108 & TOTLS index \\ \hline
    MIYS990101 & Relative partition energies derived by the Bethe approximation  \\ \hline
    MIYS990102 & Optimized relative partition energies - method A \\ \hline
    MIYS990103 & Optimized relative partition energies - method B \\ \hline
    MIYS990104 & Optimized relative partition energies - method C \\ \hline
    MIYS990105 & Optimized relative partition energies - method D \\ \hline
        ENGD860101 & Hydrophobicity index \\ \hline
    FASG890101 & Hydrophobicity index \\ \hline
    KARS160101 & Number of vertices (order of the graph) \\ \hline
    KARS160102 & Number of edges (size of the graph)\\ \hline
    KARS160103 & Total weighted degree of the graph (obtained by adding all the weights of all the vertices) \\ \hline
    KARS160104 & Weighted domination number \\ \hline
    KARS160105 & Average eccentricity \\ \hline
    KARS160106 & Radius (minimum eccentricity)\\ \hline
    KARS160107 & Diameter (maximum eccentricity)\\ \hline
    KARS160108 & Average weighted degree (total degree, divided by the number of vertices) \\ \hline
    KARS160109 & Maximum eigenvalue of the weighted Laplacian matrix of the graph \\ \hline
    KARS160110 & Minimum eigenvalue of the weighted Laplacian matrix of the graph \\ \hline
    KARS160111 & Average eigenvalue of the Laplacian matrix of the the graph \\ \hline
KARS160112 & Second smallest eigenvalue of the Laplacian matrix of the graph \\ \hline
    KARS160113 & Weighted domination number using the atomic number \\ \hline

\end{tabular}
\end{table}

\begin{table}[h!]
\begin{tabular}{|c|p{8cm}|}
\hline
    \textbf{Feature} & \textbf{Descriptor} \\ \hline
KARS160114 & Average weighted eccentricity based on the the atomic number \\ \hline
    KARS160115 & Weighted radius based on the atomic number (minimum eccentricity) \\ \hline
    KARS160116 & Weighted diameter based on the atomic number (maximum eccentricity) \\ \hline
    KARS160117 & Total weighted atomic number of the graph (obtained by summing all the atomic number of each of the vertices in the graph)\\ \hline
    KARS160118 & Average weighted atomic number or degree based on atomic number in the graph \\ \hline
    KARS160119 & Weighted maximum eigenvalue based on the atomic numbers \\ \hline
    KARS160120 & Weighted minimum eigenvalue based on the atomic numbers \\ \hline
    KARS160121 & Weighted average eigenvalue based on the atomic numbers \\ \hline
    KARS160122 & Weighted second smallest eigenvalue of the weighted Laplacian matrix \\ \hline
\end{tabular}
\end{table}

\end{document}